\documentclass[preprint,12pt]{elsarticle}
\usepackage{amssymb}
\usepackage{amsfonts}
\usepackage{amsmath}
\usepackage{graphicx}
\usepackage{epsfig}
\usepackage{multicol}
\usepackage{color}
\usepackage{xspace}
\setcitestyle{square}

\usepackage{graphicx}
\usepackage{dcolumn}
\usepackage{bm}
\usepackage{latexsym}
\usepackage{color}
\usepackage{tabularx}
\usepackage{ulem}
\usepackage{hyperref}
\usepackage{float}
\usepackage{tabularx}
\usepackage{color}
\usepackage{hyperref}
\usepackage{colortbl}

\definecolor{Gray}{gray}{0.90}
\newcolumntype{a}{>{\columncolor{Gray}}c}
\definecolor{LightCyan}{rgb}{0.88,1,1}

\newcommand\Pran{\mbox{\textit{Pr}}} 
\newcommand\Ray{\mbox{\textit{Ra}}}
\newcommand\Pen{\mbox{\textit{Pe}}}  

\newsavebox{\astrutbox}
\sbox{\astrutbox}{\rule[-5pt]{0pt}{20pt}}

\begin{document}
\title{Scalings of heat transport and energy spectra of turbulent Rayleigh-B$\acute{{\bf \mbox{e}}}$nard convection in a large-aspect-ratio box}
\vspace{-10mm}
\author[akd]{A.K.~De}
 
\author[ve]{V. Eswaran}
 
 \author[pkm]{P.K.~Mishra}
 
 \address[akd]{Department of Mechanical Engineering, Indian Institute of Technology Guwahati, Guwahati, Assam, India 781039}
 \address[ve]{Department of Mechanical Engineering, Indian Institute of Technology Hyderabad, Hyderabad, Andhra Pradesh, India 502205}
 \address[pkm]{Department of Physics, Indian Institute of Technology Guwahati, Guwahati, Assam, India 781039}
\begin{frontmatter}
\begin{abstract}
Direct Numerical Simulations of turbulent convection in a large aspect-ratio box are carried out in the range of Rayleigh number $7 \times 10^4 \le \Ray \le 2 \times 10^6$ at Prandtl number \Pran=0.71. A strong correlation between the vertical velocity and temperature is observed in the turbulent regime at almost all the length scales. Frequency spectra of all the velocities and temperature show a $-5/3$ law for a wide band of frequencies. The variances of horizontal velocities at different points in the flow yield a single power-law. Probability density functions of velocities and temperature are close to Gaussian only at higher Rayleigh numbers. The mean and variance of temperature clearly show boundary layers, surface layers and a near-homogeneous bulk region. The boundary layer thickness decreases and bulk-homogeneity is enhanced on increasing  the Rayleigh numbers. The wave number spectra of the turbulent kinetic energy exhibit kolmogorov like ($E(k)\sim k^{-5/3}$) and Bolginao-Obukhov like ($E(k)\sim k^{-11/5}$) behaviour respectively in the central and near-wall regions of the container. An approximate balance between the production due to buoyancy and the dissipation is found in the turbulent kinetic energy budget. Taylor's approximate equation of the production due to turbulent stretching and the dissipation of turbulent enstrophy is modified by the inclusion of buoyancy production in the enstrophy budget. The present results support the previously proposed $2/7$ power-law dependence of the average Nusselt number on the Rayleigh number by yielding an exponent of 0.272, but do not necessarily support the proposed classification of ``soft" and ``hard" turbulence on the basis of this exponent.
\end{abstract}

\end{frontmatter}

\section{Introduction}

Rayleigh-B$\acute{{\bf \mbox{e}}}$nard convection (RBC), in which a vertical temperature difference $(\Delta T)$ is maintained across a horizontal fluid layer, is an ideal model to understand more complex flow in the nature and the engineering flow. As $\Delta T$ is increased a systematic transition from the laminar to the turbulent states takes place in RBC. \citet{C:61} showed, for infinite horizontal layers, thermal convection starts at $\Ray_c \approx 1708$ when rolls are formed that depend on geometry and Prandtl number ($Pr$). As the Rayleigh number increases, \citet{K:87} reported transitions of stable rolls to both steady and time-dependent oscillatory flows. \citet{BC:79} described a spoke-pattern knot instability and skewed varicose instabilities which bring in considerable changes in the horizontal wavelength of the convection rolls, disappearing in the limits of small and large Prandtl numbers. Though \citet{B:67} showed that the first transition is caused by instabilities in the thermal boundary layers, it is yet to be established that turbulent convection is realized by boundary layers becoming turbulent first. A majority of low Rayleigh number studies, such as those of  \cite{K:87}, \cite{G:83}, \cite{KO:88}, \cite{Y:88}, \citet{MY1:93} and \citet{MY2:93} have reported the transition to turbulence as occurring around or below \Ray=$4 \times 10^5$, although \citet{G:82} and \citet{MO:82} have also found (laminar) periodic rolls around this Rayleigh number.   
\par
One of the interesting question in the turbulent RBC is the scaling of the energy spectra and the scaling of the Nusselt number that quantifies the efficiency of the heat transport due to the convection over the heat transport due to the conduction\cite{AGL:09}. The scaling laws of turbulent quantities have been used in the classification of different sub-regimes. Kraichnan~\cite{K:62} using mixing length theory proposed the scalings of Nusselt number as $Nu\sim Ra^{1/3}$ for large Prandtl number ($Pr>1$) and  $Nu\sim (PrRa)^{1/3}$ for small Prandtl number ($Pr<1$). Grossmann and Lohse~\cite{GL:00,GL:01,GL:02} predicted different sets of the Nusselt number scalings in the parameter space $Ra$-$Pr$. By separating the contribution in dissipation due to bulk and boundary layers, Grossmann and Lohse~\cite{GL:00} demonstrated that for the bulk-dominated convective flows generally at large $Ra$, $Nu\sim(PrRa)^{1/2}$, however, $Nu\sim(Ra)^{1/3}$ for low $Ra$. Realizing the inadequacy of classical estimates of $Nu \sim \Ray^{1/3}$ and $Nu \sim \delta_T^{-1}$ (where $\delta_T$ is the thermal boundary layer thickness), \citet{CGHKLTWZZ:89} and \citet{SS:90} obtained separate power laws for the  Nusselt number, the velocity scale normalized as Peclet number (\Pen), the temperature scale ($\Theta$) and the velocity boundary layer thickness, by assuming (a) turbulent kinetic energy dissipation in the boundary layers equals $(Nu-1)\Ray$, (b) a perfect correlation between vertical velocity and temperature exists, and (c) the velocity throughout the central region is determined by the acceleration of the vertical velocity to the top of a mixing region near the plates. Their predicted power-laws were 
\[
Nu \sim \Ray^{\frac{2}{7}},~\Pen \sim \Ray^{\frac{3}{7}},~\Theta \sim \Ray^{-\frac{1}{7}},~\delta_v \sim \Ray^{\frac{3}{7}}
\]
These theoretical estimates have served as a guideline for both measurements and numerical simulations~\cite{SC:05,WS:13,SVL:10,SVL:11,VMPP:12,PVM:14,PKCV:16}. ~\citet{PVM:14} performed numerical simulation in a box with free-slip boundary condition on the top and bottom plates and obtained the scaling $Nu\sim Ra^{0.29}$. \citet {ES:12} numerically obtained that the Nusselt number follows the scaling $Nu\sim Ra^{0.30}$ for very large $Ra$ ($10^7\le Ra \le 10^{10}$). A more extensive review of the theoretical scaling arguments has been given by \citet{K:96} and for latest overview on this issue readers are referred to the reviews by \citet{AGL:09} and \citet{CS:12} and \citet{SPGL:13}.  
\par
\citet{HCL:87} introduced the classification of ``hard" and ``soft" regimes in turbulent Rayleigh-B$\acute{{\bf \mbox{e}}}$nard convection using the power-law $Nu = A~\Ray^n$, with
\[
n=\left\{ \begin{array}{ll}
          \frac{1}{3} & \mbox{for}~~2.5\times10^5 \le \Ray \le 4\times10^7~~\mbox{(``soft")}\\
		  0.282 (\approx \frac{2}{7}) & \mbox{for}~~\Ray \ge 4\times10^7~~\mbox{(``hard")}
		  \end{array}\right.
\]
They observed that as Rayleigh number is increased beyond $\Ray_c \approx 4 \times 10^7$, the probability distribution function of the temperature fluctuations in the convective core changes from a Gaussian to an exponential distribution. They proposed the ``soft" regime is characterized by eruption of wall-attached plumes that  span the entire surface layer, while in the ``hard" regime freely moving disconnected convecting blobs of fluid exist that have broken from the large scale structures by turbulence. Furthermore, \citet{CGHKLTWZZ:89} have suggested that the distinction between ``hard" and ``soft" turbulence is an universal feature of convective turbulence and would appear to a variety of convective flows. 
\par
The theoretical model of \citet{V:95} states that temperature fluctuations are caused by instabilities in the thermal boundary layers triggered by incoming plumes; this has been later supported by the experiments of \citet{QSTX:04} and the direct numerical simulations (DNS) of \citet{K:96}. The relation between turbulent structures and the dynamics of thermal plumes is crucial to understand the nature of turbulent fluctuations and their effect on the large scale motion in turbulent convection. \citet{NSSD:00} and \citet{XLZ:02}  experimentally showed that fluctuations at different regions of the flow and the local heat transport are closely related. The large scale flow appeared as low frequency oscillations in the temperature spectra in the experiments of \citet{SWL:89} which was later replicated by \citet{AS:99}, \citet{QYT:00} and \citet{SX:01}. \citet{QSTX:04} found spatially varying vertical oscillations and almost same-frequency horizontal fluctuations, suggesting that the entire layer oscillates with almost a single frequency and amplitude. In addition to this, \citet{QT2:01} and \citet{QT:02} obtained spatial correlations which show a sharp transition from a random chaotic state to a correlated turbulent state at $\Ray_c \approx 5 \times 10^7$. An extensive discussion on this issue can be found in the review articles written by \citet{AGL:09} and \citet{Sig:94}.
\par
Numerical studies of turbulent Rayleigh-B$\acute{{\bf \mbox{e}}}$nard convection are comparatively few, mainly due to the  severe computational constraints in high-Rayleigh number flows. Earlier two-dimensional work by \citet{DW2:65} and \citet{LS:71} over-predicted the heat transfer rates and the transitional Rayleigh number. Early three-dimensional simulations were either limited to studying the laminar-turbulent transition, as in \cite{LS:71} and \cite{OYCS:76}, or lower Rayleigh number turbulent flow, as in \cite{L:76}. However, many complex phenomena were successfully observed in numerical studies. For instance, \citet{G:82} observed the `skewed varicose' instability at \Ray=$4 \times 10^3$. \citet{WG:98} explained the importance of pressure transport in turbulent convection, using DNS. However, although their highest Rayleigh number (\Ray=$10^9$) was in the ``hard" turbulence regime, the grid resolution of their simulations was limited even compared to, say, the DNS of \citet{K:96} where the highest Rayleigh number was two orders smaller. 
\par
\citet{PCG:02} performed DNS for low Rayleigh number $(7 \times 10^3 \leq \Ray \leq 10^5)$ cases, while for their highest two Rayleigh numbers ($10^6$ and $10^8$) they used large-eddy simulations (LES). \citet{Kerr:85} had earlier shown that, in order to resolve dissipation statistics accurately, at least one decade of the calculation must be dedicated to the viscous regime. \citet{KMM:87} in their channel flow simulations and \citet{G:83} in turbulent convection supported this criterion compared to the previous orthodoxy that for DNS grid resolution should be made to resolve the Kolmogorov length scale. The DNS of \citet{K:96}, however, were spatially well-resolved even from this consideration. \citet{KH:99} using RANS-LES studied the scaling of Nusselt number for $Ra=10^6-10^9$ and obtained the exponent consistent with the experimental observation. \citet{KH:06} used the hybrid RANS/LES merging scheme to capture the dynamics and characteristics of the heat transport accurately near the boundary layer region as well as in the bulk and demonstarted that $Nu$ scaling exponent as well as the vertical spatial variation of the mean temperature and temperature fluctuation match quite well with the DNS~\cite{K:96,W:94} and experimental observations~\cite{NSSD:00}. \citet{ZG:15} based upon the RANS-LES studies obtained that the Nusselt number scaling exponent is close to $1/4$ for $Ra\sim10^7$, however, at higher Rayleigh number ($Ra\sim10^8$) scaling exponent was more closer to the $1/3$. Silano \textit{et al.}~\cite{SSV:10} numerically studied the Nusselt number for a wide range of Prandtl number ($Pr=10^{-1}$-$10^4$) and reported that the Nusselt number exponent is close to $2/7$ for $Pr=1$ and $0.31$ for $Pr=10^3$. \citet{SPA:16} performed the direct numerical simulation in a cylinder container of aspect-ratio 6.3 for $Pr=6.7$ and reported a coherent structure also known as the mean-wind in the turbulent regime.
\par
Overall, numerical simulations have been able to only confirm a few transitions in a Rayleigh number range significantly lower than what  \citet{HCL:87} experimentally achieved. Given the constraints imposed by the computational cost of high-Rayleigh number simulations, an important motivation for this study is to check the possibility that ``hard" turbulence may be attained at much lower Rayleigh numbers in large aspect-ratio simulations compared to the small-aspect ratio all-no-slip boundaries experiments\cite{AGL:09}, as was seen by \citet{WL:92}, or computationally by \citet{WDRC:91}. Such simulations will not only test the $Nu \sim \Ray^{2/7}$ scaling but they can also give new insights into turbulent convection at high Rayleigh number. 

\section{Mathematical and numerical description}\label{sec:mathnum}

\subsection{Boussinesq equations}
The governing equations for an incompressible buoyancy driven flow (using the Boussinesq approximation) are the continuity,  momentum and the energy equations, written in normalized form as follows 
\begin{eqnarray}
\label{nce}
\frac{\partial u_j}{\partial x_j} &=& 0 \\
\label{nme}
\frac{\partial u_i}{\partial t}+ \frac{\partial(u_i u_j)}{\partial x_j} &=& -\frac{\partial p}{\partial x_i} +\sqrt {\frac{\Pran}{\Ray}} \frac{\partial^2 u_i}{\partial x_j \partial x_j}+ \theta \delta_{iy} \\
\label{nee}
\frac{\partial \theta}{\partial t}+ \frac{\partial(u_j \theta)}{\partial x_j} &=& \frac{1}{\sqrt{\Ray \Pran}} \frac{\partial^2 \theta}
{\partial x_j \partial x_j}
\end{eqnarray}
\noindent
where \Ray $(\equiv g\beta\Delta T H^3/\nu \alpha)$ and \Pran $(\equiv \nu/\alpha)$ are respectively the Rayleigh and Prandtl numbers. Normalization of the equations is done using the buoyancy velocity scale $U=\sqrt{g \beta \Delta T H}$, the height of the box, $H$, and the temperature difference between the top and bottom boundaries, $\Delta T$. All the time scales present in the system are non-dimesionalized using the free-fall time $t_f=H/\sqrt{\beta g \Delta T H}$. 

\subsection{Numerical technique}
\begin{figure}
\centerline{\includegraphics[scale=0.6]{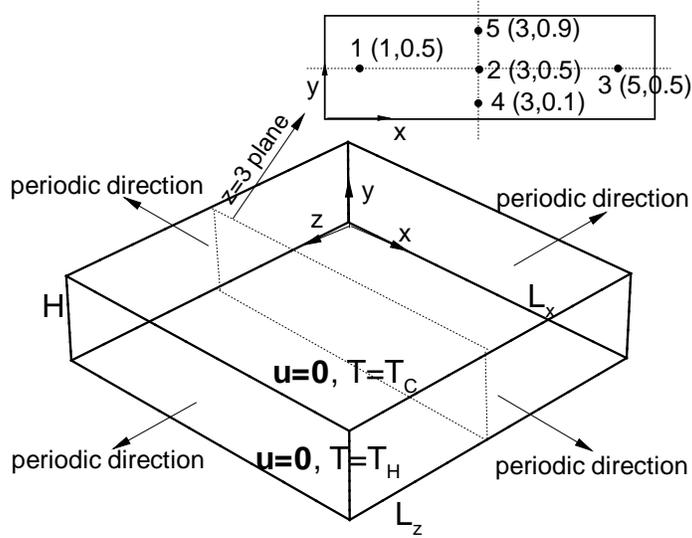}}
\caption{Geometrical details with signal points.}\label{fig:geo}
\end{figure}

The set of coupled governing Eqs. \eqref{nce}-\eqref{nee} are solved by a semi-explicit method. A staggered grid arrangement for the variables is used, with temperature defined at the cell center. The solution method uses a two-step predictor-corrector algorithm wherein the provisional (i.e., predicted) velocity field is computed in the first step, using previous time-level $(n)$ pressure values, by the $2^{nd}$-order accurate Adams-Bashforth Crank-Nicolson time integration scheme; in the second step these provisional values are then corrected to obtain a divergence-free velocity field at the new time-level $(n+1)$. The latter requires the solution of a Poisson equation for the pressure correction. The temperature values are then advanced using this divergence-free velocity field using the same time integration method. The $4^{th}$-order central explicit scheme of \citet{T:93} with enhanced spectral resolution is used for the convective discretization. This scheme, with a seven point stencil, does not introduce higher order numerical dissipation and, optimized for a better spectral resolution property, has the following difference formula:
\[
\phi^{\prime}_i=\frac{1}{h}[0.02651995~(\phi_{i+3}-\phi_{i-3})-0.18941314~(\phi_{i+2}-\phi_{i-2})+0.79926643~(\phi_{i+1}-\phi_{i-1})]
\]
where ${\phi_i}^{\prime}$ is the spatial derivative of $\phi$ at the grid point $i$, and $\phi_{i+1}$ etc, are the values of $\phi$ at the equally-spaced neighboring points in the derivative direction. Essentially the same formula can be used on non-uniform rectangular grids by introducing a simple geometric scaling coefficient. Diffusion terms are discretized by the $2^{nd}$-order central difference scheme. The geometric details along with the boundary conditions are shown in Fig. \ref{fig:geo}. The grids are uniform along the periodic-horizontal directions, while in the vertical direction grid-refinement is used near the walls. In all the calculations $211 \times 211 \times 67$ grids are used in $x,z$ and $y$ directions, with 
\[
\Delta x=\Delta z=2.8571 \times 10^{-2},~~\Delta y_{max}= 2.8571 \times 10^{-2},~~
\Delta y_{min}= 3.2866 \times 10^{-3}
\] 
The grid-size and the minimum spacing are chosen based on the estimates of the Kolmogorov length scale $(\eta)$ by \citet{K:96} in the present range of Rayleigh numbers. From the lowest to the highest Rayleigh numbers, $\Delta x/\eta,\Delta y/\eta,\Delta z/\eta$ vary approximately from 1/2 to 2 in the bulk while near the walls $\Delta x/\eta,\Delta z/\eta$ vary from 3/4 to 2 and $\Delta y/\eta$ from 1/10 to 1/4. The time increment was fixed at $\Delta t=10^{-3}$ (in units of free-fall time) for all the calculations; no numerical instabilities were observed at this value. The solver was parallelized on an MPI-Fortran platform. Once the simulated flow became fully turbulent, at least 10-15 large eddy turn-over times were computed so as to have sufficient sampling time for the temporal statistics. Planar averages were taken only after checking for the statistical stationarity of the simulated turbulent data. 
\begin{figure}[!htp]
\centerline{\includegraphics[scale=0.5]{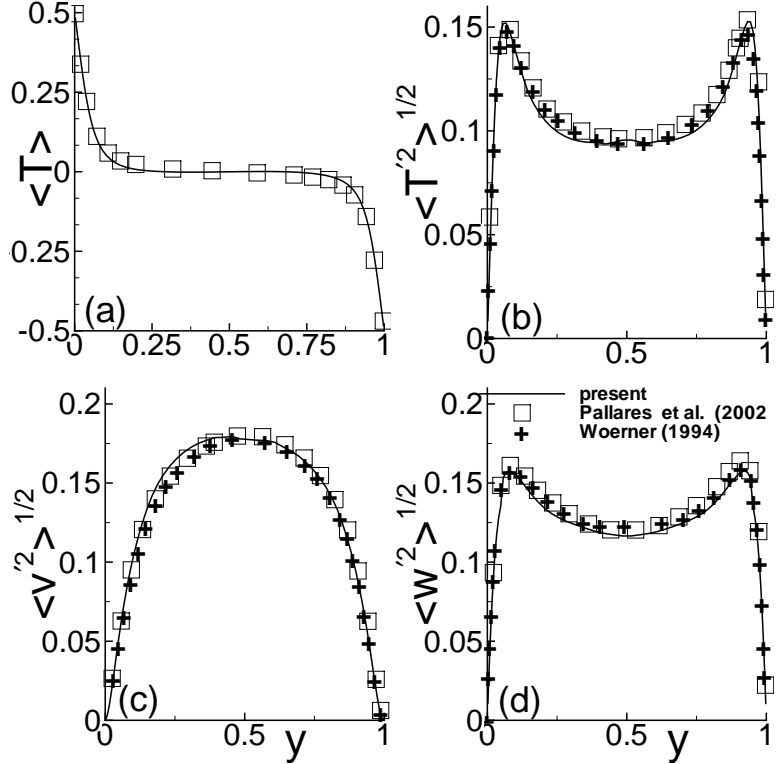}}
\caption{The vertical variations of (a) mean temperature, (b) root mean square fluctuation of temperature and (c,d) root mean square velocity fluctuations for $\mathrm Ra=6.3\times 10^5$ and $Pr=0.71$. \textemdash: data from our simulation; $\square$: simulation data taken from Pallares {\it et al.} \cite{PCG:02};  $+$: data taken from Woerner \cite{W:94}.}\label{fig:comp}
\end{figure}
\par
The parallelized numerical code was first used to compute a turbulent convection case in a 6:6:1 aspect ratio box at \Ray$=6.3 \times 10^5$, so that the results could be compared with the DNS results of \citet{W:94}, \citet{K:96} and \citet{PCG:02}. 
\begin{table}
\begin{center}
\scalebox{0.8}{
\begin{tabular}{lccccc} \hline
 &  \Ray & Aspect-ratio & $N_x \times N_z \times N_y$ & $\Delta x_{min},\Delta y_{min}$ & 
 $<Nu>$ \\
\citet{W:94} & $6.3 \times 10^5$ & 7.9:7.9:1 & $200 \times 200 \times 49$ & 0.0369,~0.005 & 7.27 \\
\citet{K:96} & $5 \times 10^5$ & 6:6:1 & $96 \times 96 \times 48$ & 0.0625,~0.002 & 7.46 \\ 
\citet{PCG:02} & $6.3 \times 10^5$ & 6:6:1 & $81 \times 81 \times 61$ & 0.0302,~0.004 & 7.6 \\
Present & $6.3 \times 10^5$ & 6:6:1 & $91 \times 91 \times 61$ & 0.0667,~0.0036 & 7.415 \\    \hline
\end{tabular}}
\caption{Computational features of the comparison test.}
\label{compar}
\end{center}
\end{table}
In the present simulation, all the computational features (see table \ref{compar}) including the boundary conditions were kept as close as possible to the above references. Figure \ref{fig:comp} shows close agreement of the mean temperature $(a)$ and the root mean square of temperature $(b)$ and velocities $(c,d)$ with the earlier simulations. Table \ref{compar} also shows the computed average Nusselt number ($<Nu>$) is within $\pm 2\%$ of the reported values. This comparison shows a satisfactory performance of the present code compared with the earlier simulations.

\section{Results}\label{sec:res}
Detailed discussions on the temporal and spatial statistics, mainly for three Rayleigh numbers that span the present simulation range, now follow separately. 

\subsection{Temporal statistics}  
\begin{figure}[!htb]
\centerline{\includegraphics[scale=0.60]{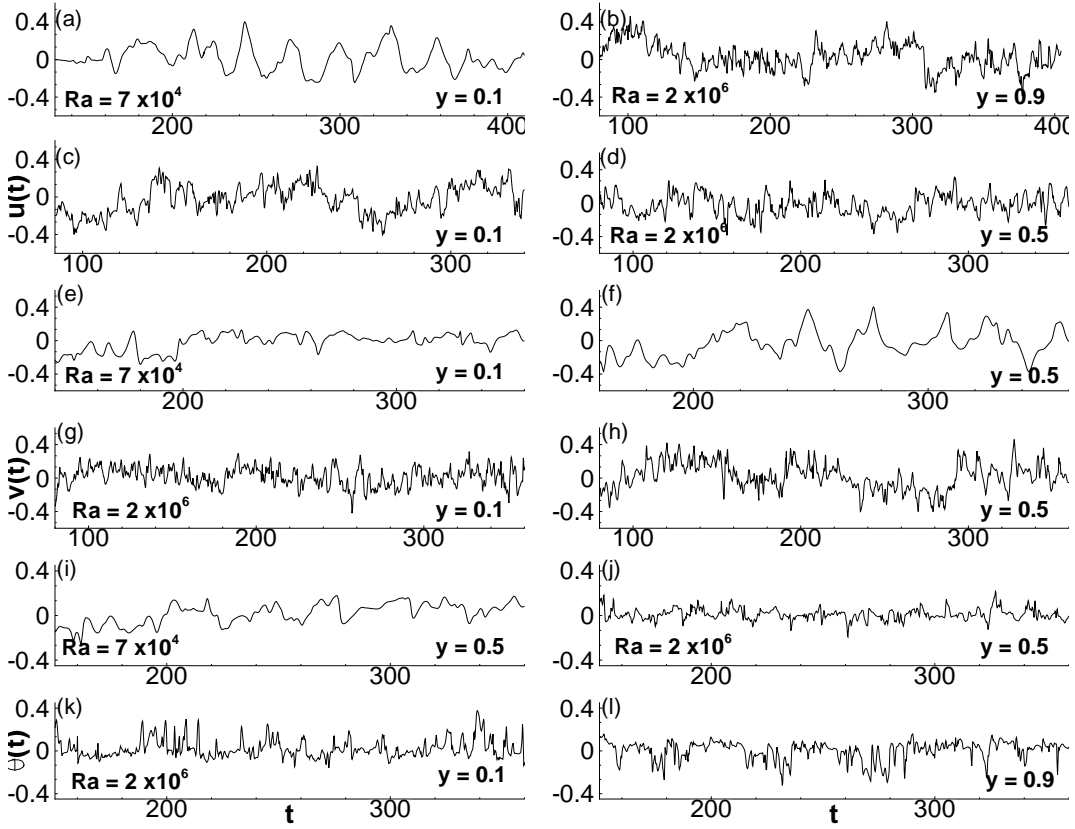}}
\caption{$u$ $(a-d)$, $v$ $(e-h)$ velocity and temperature $(i-l)$ signals; each row of figures correspond to the same Rayleigh number, unless specified otherwise.}\label{fig:signal}
\end{figure}
\par
In computing the statistics of the flow, special attention is paid to five points on the mid-$z$ plane $(z=3)$, shown in Fig. \ref{fig:geo}, chosen to characterize the flow near the walls and at the bulk. Figure \ref{fig:signal} shows a few selective time-series of the velocities and temperature spanning 400, 320 and 350 time units which correspond approximately to 12, 25 and 40 integral time-scales, respectively, for \Ray$=7 \times 10^4,5 \times 10^5$ and $2 \times 10^6$.
\par
In the time-series, the effect of Rayleigh number is clearly seen in the change from the relatively smooth signals at \Ray$=7 \times 10^4$ to the jagged signals at \Ray$=2 \times 10^6$, with the advent of smaller time-scales. The vertical velocity and temperature are clearly correlated near the walls, suggesting the  buoyancy force results in the correlation $\overline{v^{\prime}\theta^{\prime}}$ in turbulent convection that sustain fluctuations. While the horizontal velocity components are not obviously correlated, peaks (positive or negative) of one horizontal component are simultaneous to high values of the other component near the walls (see figures $b$ and $c$) which suggests perturbation of horizontal velocities in the boundary layers due to strong upwash/downwash of plumes. Thus, the overall picture suggests that the buoyancy force directly acts as the motive force on the vertical component of velocity and indirectly forces the other components through the action of the plumes and their upwash and downwash.
\par
Horizontal movements are stronger near the plates, leading to larger oscillations there, compared to the bulk region (see figures $b,c$ and $d$). At the highest Rayleigh number, $u(t)$ shows sharp peaks and valleys as a result of violent interactions of adjacent plumes, and sustained positive and negative oscillations are seen near the walls (figures $b$ and $c$) for much longer time intervals than the estimated integral time scale. This hints that the effects of stronger plumes, both hot and cold, last over time scales significantly higher than what autocorrelation coefficients seem to indicate. 
\par
The vertical velocity shows higher amplitudes of oscillation in the bulk compared to the
near-boundary regions at all Rayleigh numbers (see figures $e,f$ and $g,h$). This is because the plumes, that form at the walls as a result of boundary layer instability, accelerate towards and reach the bulk from both directions, causing high fluctuations. However, temperature fluctuations reduce to a white-noise-like pattern (see figure $j$) with even smaller amplitudes at the bulk region, owing to a near-homogeneous state there.
\par
On the other hand, points near the plates show predominantly hot ($y=0.1$, figure $k$) and cold ($y=0.9$, figure $l$) fluctuations about the zero mean which correspond to eruptions of plumes due to boundary layer instabilities. These time-series closely resemble those in the experiment of \cite{SWL:89} for a cylindrical cell of aspect ratio 1 at \Ray$=5 \times 10^{10}$. This suggests features of turbulence seen in smaller aspect-ratio cells may indeed appear at a much lower Rayleigh number for a larger aspect-ratio box due to the absence of the restrictive constraints of the lateral walls.               

\subsubsection{Frequency spectra}
\begin{figure}[!htp]
\centerline{\includegraphics[scale=0.75]{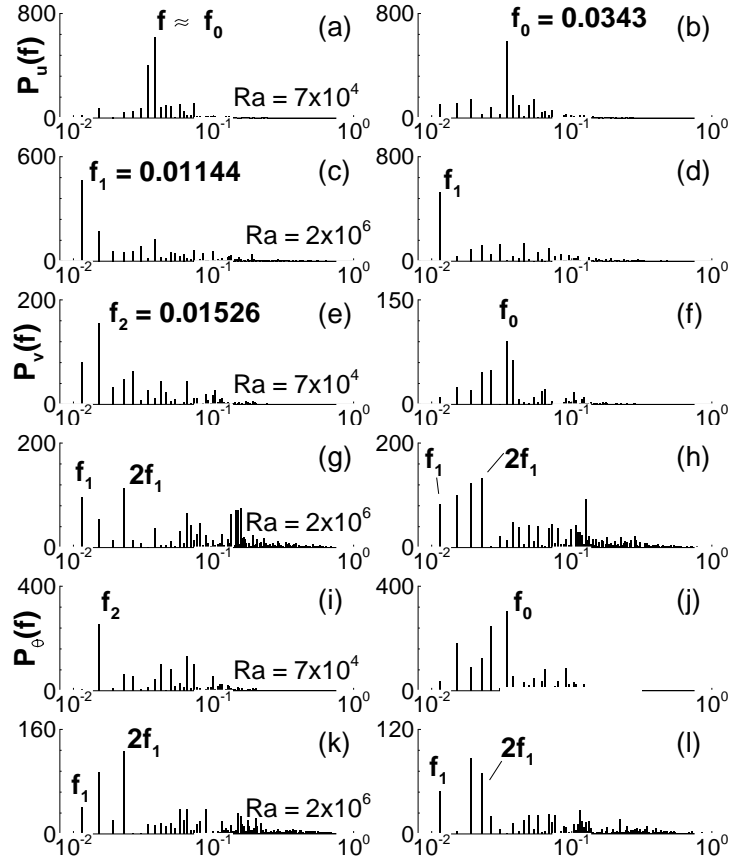}}
\caption{Dominant peaks in the frequency spectra; two figures in a row correspond to the same Rayleigh number. Left and right panels are respectively near the hot plate (at $y=0.1$) and near the cold plate (at $y=0.9$). }\label{fig:frpeak}
\end{figure}
\par
The frequency spectra, computed by taking the FFTs of the appropriate auto-correlation functions, are shown in Fig. \ref{fig:frpeak}$(a-l)$ with distinct dominant frequencies identified as $f_0,~f_1,~f_2$. 
\par
Both the horizontal velocity components oscillate with the same frequency, $f_0=0.0343, 0.01907$ and $f_1=0.01144$ for $\Ray=7\times10^4,~5\times10^5$ and $2\times10^6$, respectively, with several harmonics appearing at the bulk at increasing Rayleigh numbers. The vertical velocity and temperature frequency plots show a strong correlation at any specific point. They also show at all points roughly the same dominant frequencies which,  however, are not always identical to the dominant frequencies for the horizontal velocities. At the lowest Rayleigh number, points near the plates $(y=0.1,0.9)$ show peaks at $f_2=0.01526$ and $f_0=0.0343$ (same as in $u$ and $w$), respectively. However, the $f_2$ frequency (see Fig. \ref{fig:frpeak}$e)$ is not present in the horizontal velocity, and seemingly corresponds to the diffusion time scale (which is $\sqrt{\Ray~\Pran}$ times the buoyancy time scale associated with the $v-\theta$ correlations). At the intermediate and the highest Rayleigh numbers, these same features are seen with harmonics of the primary frequency being observed. It is evident that with increase in Rayleigh number the vertical oscillations are more nearly aligned with the temperature fluctuations owing to the stronger buoyancy force.
\begin{figure}[!ht]
\centerline{\includegraphics[scale=0.6]{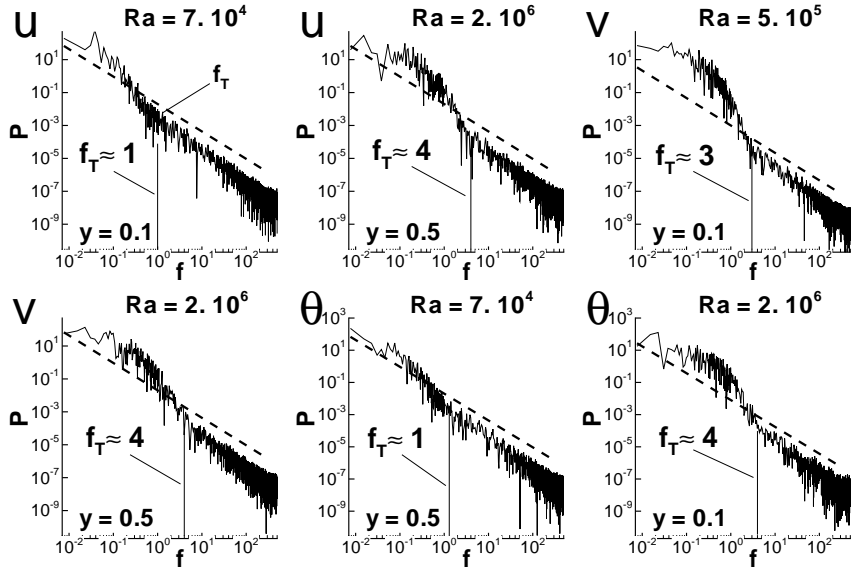}}
\caption{Frequency spectra for the velocities and temperature showing threshold frequency $f_T$ for the power-law variation.}\label{fig:frspc}
\end{figure}
\par
A wider band of excitation frequencies is realized with increase in Rayleigh number. A few clearly dominant peaks of $u$ and $w$ are visible in the near-wall layers, while at the bulk a larger number of dominant peaks appear. This feature reverses for the $v$-velocity. As Rayleigh number increases, no clear difference in the dominant frequencies of the horizontal and vertical fluctuations can be identified. Thus, it emerges that oscillating frequencies of all the velocities and temperature are nearly same, and more so as Rayleigh number increases.       
\par
The frequency spectra, plotted on a log-log scale, are shown in Fig. \ref{fig:frspc} where a power law 
\[
P(f)=\left ( \frac{f}{f_c} \right )^{-s}
\]
has been fitted (with $f_c$ being a fitting parameter).

It was found that the frequency spectra of all velocities and temperature align with a $f^{-5/3}$ law in a wide band of frequencies. The threshold frequencies $f_T$ (which is very nearly the same for the velocities and temperature) beyond which the power-law closely represents the data, increases from $f_T \approx 1$ at $\Ray=7\times10^4$ to $f_T \approx 4$ at $\Ray=2\times10^6$, suggesting the dominance of low frequency modes with increase in Rayleigh numbers. A similar distribution of energy modes in the intermediate frequency range was also reported by \citet{SWL:89} in their unit aspect-ratio cylindrical cell experiment in the range $10^8 < \Ray < 10^{11}$ but with $s \approx 1.4$. This difference $(s=5/3$ vs 1.4) in the exponent weakens the analogy between lower Rayleigh number flow in a large aspect-ratio box and very high Rayleigh number flow in a small aspect-ratio box. The calculated spectra of $\theta$ is nearly same as that of the velocities suggesting that the temporal oscillations of velocities and temperature have close correlation even at the smaller time-scales. 

\subsubsection{Probability distribution function of velocity and temperature fluctuations}
\par
A transition of velocity and temperature fluctuation probability density functions (pdfs) from a Gaussian distribution to an exponential distribution has been reported by \citet{SWL:89} and \citet{QSTX:04} and proposed as the distinguishing feature between ``soft" and ``hard" turbulence. However, observations from numerous experimental and numerical studies differ based on the Rayleigh number range and aspect ratios studied, and a unified view has not yet emerged. At smaller aspect-ratios $(A=1)$, while \citet{SWL:89} reported the transition as Rayleigh number increased from \Ray=$10^6$ (``soft") to \Ray$=4 \times 10^{10}$ (``hard"), \citet{DWR:90}, \citet{S:90} and \citet{SG:91} obtained exponential distributions in the range $10^6 < \Ray < 2 \times 10^8$. In contrast to this \citet{QSTX:04} showed a clean Gaussian profile even at \Ray$=3.7 \times 10^9$. On the other hand, at higher aspect-ratios, \citet{SBM:89} and \citet{KW:90} reported exponential distributions at \Ray$=6.5 \times 10^6,~A=2 \sqrt{2}$ and \Ray=$5 \times 10^5,~A=6$, while \citet{CD:93} showed mixed Gaussian and exponential distributions at moderate Rayleigh numbers ($2.5 \times 10^5 <$ \Ray $< 6.3 \times 10^5$) for aspect-ratio $A \approx 5$. 
\par
\begin{figure}[!ht]
\centerline{\includegraphics[scale=0.7]{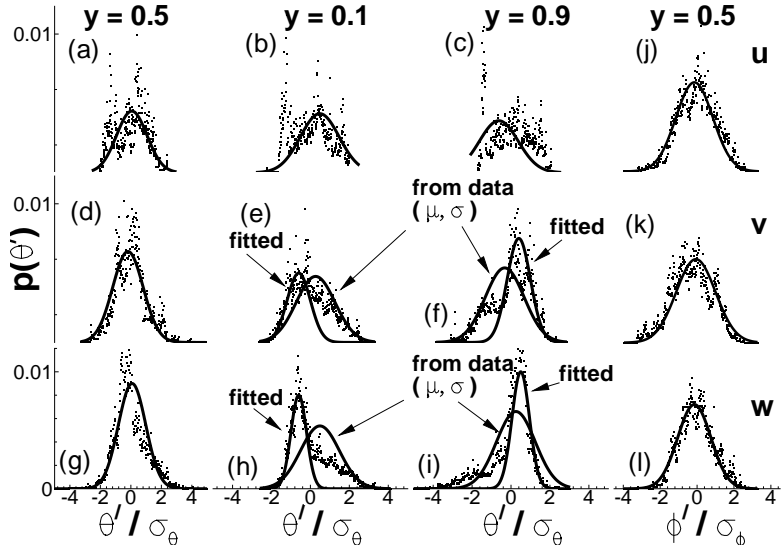}}
\caption{Temperature $(a-i)$ and velocity $(j-l)$ fluctuation pdfs, Gaussian profiles -- with the same computed mean and standard deviation (``from data") and a visual fit (``fitted") to identify the tails shown separately. \Ray$=7 \times 10^4~(a-c),~\Ray=5 \times 10^5~(d-f),$ and $\Ray=2 \times 10^6~(g-l)$.}\label{fig:pdf}
\end{figure}

In the present study, the pdfs are computed using 500 fluctuation value windows which are normalized by the root mean square, shown in Fig. \ref{fig:pdf}. 
We have calculated the Gaussian distribution, labeled ``from data", using the same mean $(\mu)$ and the standard deviation $(\sigma)$ as the simulation data, while keeping areas under the actual pdfs and the Gaussian profiles equal.  
\[
y=A \exp \left [ -\frac{1}{2} \left ( \frac{x-\mu}{\sigma} \right )^2 \right ],~A=\frac{\mbox{Area under the actual pdfs}}{\sigma \sqrt{2 \pi}}
\] 
At \Ray=$5 \times 10^5$ and $2 \times 10^6$, a second ``fitted" Gaussian distribution is visually fitted to identify the long positive or negative tails that appear in the pdfs near the walls.
\par 
At the lowest Rayleigh number the data do not fit the theoretical Gaussian curve showing a long plateau for the entire fluctuation range, which drops almost vertically at the highest fluctuation levels $(\left | \theta^{\prime}/\sigma_{\theta} \right | \approx 2)$. With increase in Rayleigh number, while the Gaussian-fit improves near the bulk, long positive and negative tails are evident in the flow near the walls. Velocity fluctuation pdfs (Fig. \ref{fig:pdf}$j-l$) are seen to be almost the same as the $\theta$ pdf and are seen to be Gaussian only at \Ray$=2 \times 10^6$. \citet{QSTX:04} found velocity pdfs with two peaks due to two levels of oscillations at $3.7 \times 10^9$ in an unity aspect-ratio cell. This latter result, however, cannot be compared with the present simulations because of the presence of lateral side walls and the Rayleigh number range of three decades higher of the earlier study. 
\par
The present results strongly support the argument of \citet{CD:93} that at moderate Rayleigh numbers, a mixed Gaussian and exponential distribution slowly changes to a pure Gaussian in the central region as the Rayleigh number increases. Thus, the classification of ``hard" and ``soft" turbulence in Rayleigh-B$\acute{{\bf \mbox{e}}}$nard convection based on the purportedly abrupt change in the nature of the temperature pdf is not supported by the present results, despite the higher aspect-ratio being used.

\subsubsection{Rayleigh number dependence of velocity and temperature variances}
\begin{figure}[!htb]
\centerline{\includegraphics[scale=0.5]{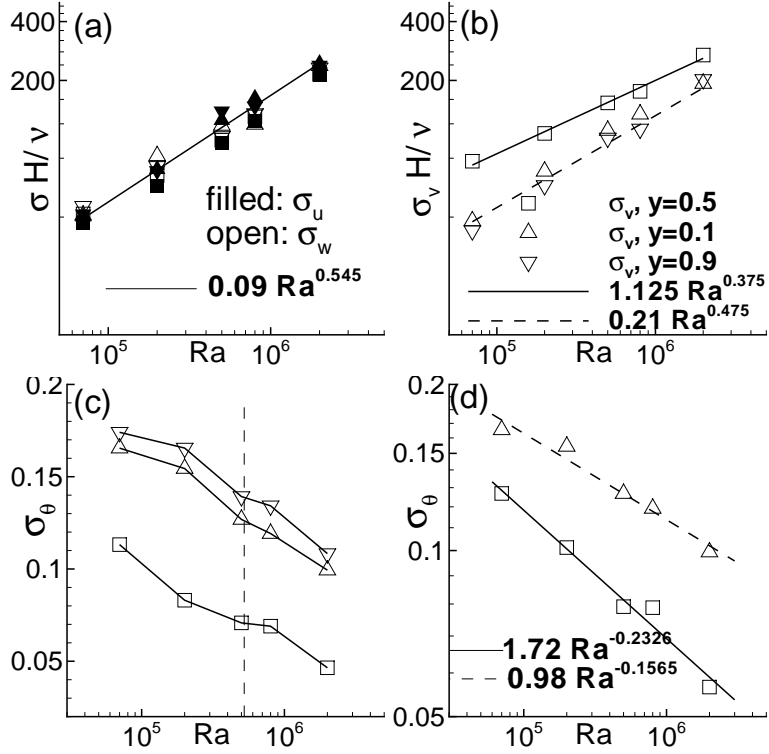}}
\caption{Rayleigh number dependence of r.m.s of the velocities (normalized to the Peclet number) and temperature; symbols corresponding to different heights are same as in $(b)$.}\label{fig:rmsRa}
\end{figure}

\par
The dependence of the normalized velocity and temperature r.m.s values on the Rayleigh number are shown in Fig. \ref{fig:rmsRa}. The points from the bulk and near-wall regions fall on a single curve (Fig. \ref{fig:rmsRa}$a$) for the horizontal oscillation. On the other hand, the higher Rayleigh number exponent (0.475) at $y=0.1,0.9$ compared to the bulk (0.375) for $\sigma_v$ (Fig. \ref{fig:rmsRa}$b$) shows that with increase in Rayleigh number that the boundary layers are perturbed more. 

The even higher exponent 0.545, seen for $\sigma_u,\sigma_z$ in Fig. \ref{fig:rmsRa}$a$, indicates that the dominant mode of instabilities are the boundary layers which primarily originate due to horizontal velocity variation, supporting the theoretical model of \citet{V:95}. \citet{QSTX:04} found the exponents as 0.65 for $\sigma_u,\sigma_z$ and 0.55 for $\sigma_v$ in the range $10^7 < \Ray < 10^9$ in a $A=1$ container.  This difference with the present results possibly indicates that lateral confinement has a more repressive effect on vertical oscillations at lower Rayleigh numbers, yielding a higher exponent.     
\par
The variation of $\sigma_{\theta}$ (Fig. \ref{fig:rmsRa}$c$) identifies a narrow transition zone at \Ray $\approx 5 \times 10^5$, analogous to the transition of ``second soft turbulence'' to ``hard turbulence" state at \Ray$=4 \times 10^7$ reported by \citet{SWL:89}. The characteristics of these two regimes in the present case are bursts of hot and cold fluids and two-level switching, respectively, seen in Figs. \ref{fig:signal}$(j,k,l)$ which is in agreement with the above literature. However, due to the non-availability of more Rayleigh number data, this ``classification" is only based on observations of time signals and the abrupt change in the $\sigma_{\theta}-\Ray$ curves. The bulk region yields a higher decay exponent (-0.2326) compared to the near wall regions (-0.1565) (see Fig. \ref{fig:rmsRa}$d$) as mixing at the bulk homogenizes the thermal field more strongly for higher Rayleigh numbers while the near-wall regions continue to be significantly dominated by plumes (which are inherently non-homogeneous). 

\subsection{Planar Statistics}

\noindent
The flow being homogeneous in the horizontal directions, the values of a given variable at two different points on a particular horizontal plane are assumed to be different realizations of the same random variable. Thus, as the flow field is statistically stationary, the time-averages (over a period of several integral time scales) of the instantaneous horizontal plane-averaged quantities may be thought of as ensemble-averages. Since there are no physical reasons for directional biases in the horizontal planes, any discrepancy of computed statistical quantities in $x$ and $z$ directions can be assumed to be the result of inadequate sampling time, more likely to occur for smaller Rayleigh numbers, which have larger integral time-scales. Planar statistics are computed over a  period of 2-4 integral time scales at all the Rayleigh numbers, and the difference in the $x$ and $z$ statistics are monitored, as a check of consistency.

\subsubsection{Mean temperature and variances}
\begin{figure}[!htb]
\centerline{\includegraphics[scale=0.7]{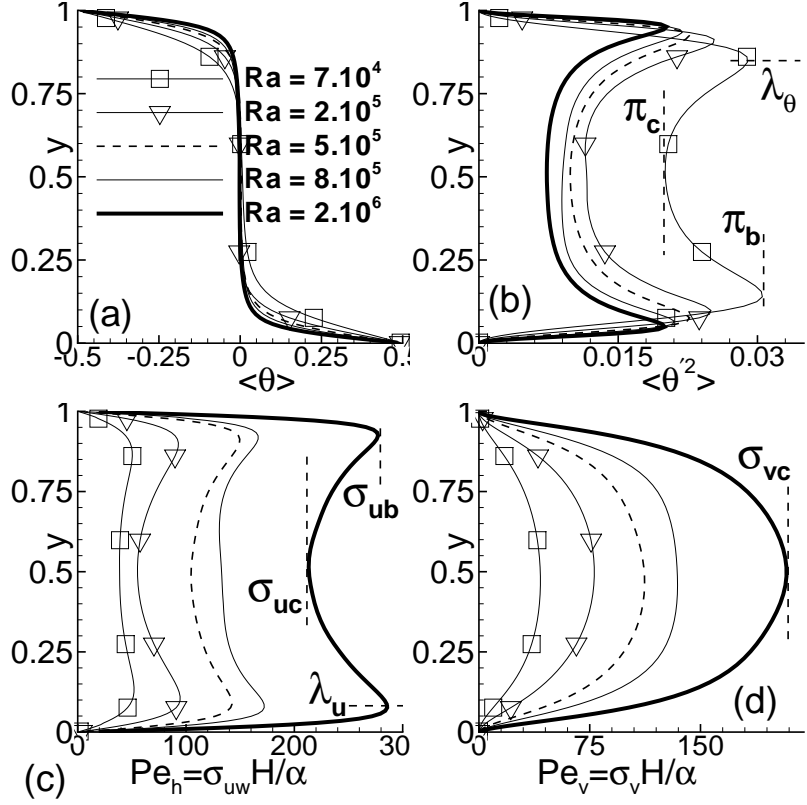}}
\caption{Vertical profiles of $(a)$ mean temperature, $(b)$ temperature variance, $(c)$ horizontal velocity variance and $(d)$ vertical velocity variance; labels are same as in $(a)$.}\label{fig:meanvar}
\end{figure}
The vertical profiles of the mean and variance of temperature, shown in Fig. \ref{fig:meanvar}$(a-b)$, identify different characteristic regions in the domain. Both $<\theta>$ and $<{\theta^{\prime}}^2>$ indicate thermal boundary layers, having regions of sharp gradients that become thinner with increase in Rayleigh number. The gradients gradually drop to reach an asymptotic value in the bulk region through the surface layer where the curves gradually falls from their extrema. 

At $\Ray=2\times 10^6$, an almost constant value of $<\theta>$ and $<{\theta^{\prime}}^2>$ in the bulk again suggests that the mixing of hot and cold fluids leads to enhanced homogeneity there. The variance curves shift lower as Rayleigh number increases due to an increase in small scale temperature fluctuations, also observed in the time-series data. 
\par
The velocity variances ($\sigma_{uw}=\sqrt{\sigma_u^2+\sigma_w^2}$ and $\sigma_v$), normalized to the Peclet number $\Pen (\sigma H/\alpha)=\sqrt{\Ray.\Pran} ~\sigma^{*}$ where $\sigma$ and $\sigma^*$ are the dimensional and non-dimensional variances, respectively, are shown in Fig. \ref{fig:meanvar}$(c-d)$. 
While $Pe_h$, in agreement to the classical turbulent boundary layer profile of \citet{KMM:87}, peaks near the solid surfaces and gradually reduces to a central value, $Pe_v$ continues to increase beyond the point of maximum $\sigma_{uw}$ which is contrary to the observation made by them. The gradient of $Pe_v$ near the walls is less steep than $Pe_h$ but remains positive beyond the boundary layer to form a rounded profile near the center. Velocity and temperature variances are in qualitative agreement with the experimental results of \citet{DW:67} and the numerical predictions of \citet{MR:90} and \citet{K:96}.   
\par
\begin{figure}[!htb]
\centerline{\includegraphics[scale=0.6]{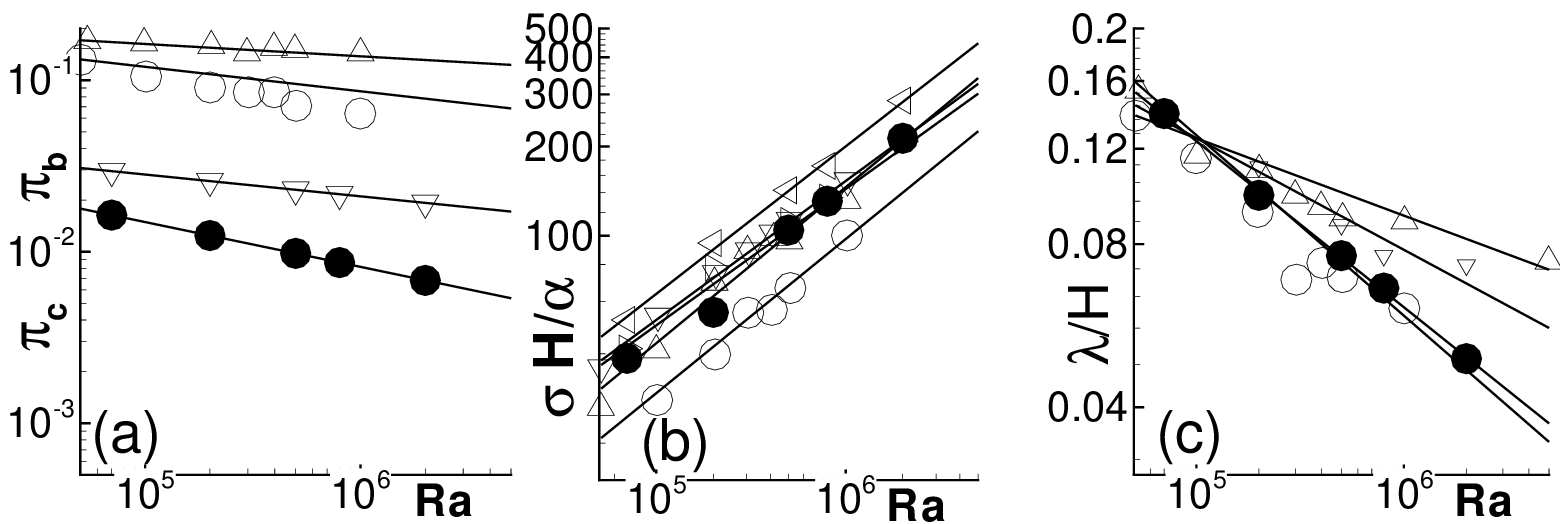}}
\caption{Power-laws of the velocity, temperature variance and boundary layer scales shown in Fig. \ref{fig:meanvar}; $(a)$ \citet{K:96}: $\circ~\pi_c(0.62Ra^{-1/7})$, $\bigtriangleup~\pi_b(0.37Ra^{-1/14})$; Present: $\bullet~\pi_c(0.3Ra^{-0.2615})$, $\bigtriangledown~\pi_b(0.12Ra^{-0.1271})$, $(b)$ \citet{K:96}: $\circ~\sigma_{uc}(0.074Ra^{0.52})$, $\bigtriangleup~\sigma_{ub}$ and $\bigtriangledown~\sigma_{vc}(0.25Ra^{0.46})$; Present:  $\bullet~\sigma_{uc}(0.1Ra^{0.5264})$, $\lhd~\sigma_{ub}(0.21Ra^{0.4974})$, $\rhd~\sigma_{vc}(0.24Ra^{0.4686})$, $(c)$ \citet{CGHKLTWZZ:89} and \citet{K:96}:  $\circ~\lambda_{\theta}(5.9Ra^{-1/3})$, $\bigtriangleup~\lambda_u(0.65Ra^{-1/7})$; Present: $\bullet~\lambda_{\theta}(4.24Ra^{-0.3068})$, $\bigtriangledown~\lambda_u(1.35Ra^{-0.2063})$}\label{fig:velTscale}
\end{figure}

The scales, shown in Fig. \ref{fig:meanvar}$(a-d)$, serve as estimates for both velocity $(\sigma_{uc},\sigma_{ub},\sigma_{vc})$ and temperature ($\pi_b,\pi_c$) and their boundary layers ($\lambda_{\theta},\lambda_u$). The assumed power law for these scales are $\pi_c \sim \Ray^{\alpha_c},\pi_b \sim \Ray^{\alpha_b},\sigma_{uc} \sim \Ray^{\beta_1},\sigma_{ub} \sim \Ray^{\beta_2},\sigma_{vc} \sim \Ray^{\beta_3},\lambda_{\theta} \sim \Ray^{\gamma_T}$ and $\lambda_u \sim \Ray^{\gamma_u}$ which are shown in Fig. \ref{fig:velTscale}$(a-c)$. 
\par
The variance $<{\theta^{\prime}}^2>$ is found to be significantly less steep (see Fig. \ref{fig:velTscale}$a$) near the walls compared to the bulk ($|\alpha_b|<|\alpha_c|$) which is consistent with \citet{WL:92}. However, both $\pi_c$ and $\pi_b$ are significantly less compared to the reported predictions of \cite{CGHKLTWZZ:89},\cite{WL:92} and \cite{K:96}. The estimates for $\alpha_c$ are -1/9 by classical theories, -1/7 by the hard turbulence scaling of \cite{CGHKLTWZZ:89}, analytical arguments of \cite{SS:90} and the DNS of \cite{K:96} compared to -0.2615 in the present calculation which is closest, among these, to the large-aspect-ratio result (-0.20) of \cite{WL:92}. The present exponent $\alpha_b$ is -0.1271 compared to $-1/14$ of \cite{K:96}. Thus the present scaling estimates of $<{\theta^{\prime}}^2>$ only roughly agrees with the previously reported data. For the velocity variances, the scaling exponents $\beta_1,\beta_2,\beta_3$, plotted in Fig. \ref{fig:velTscale}$(b)$, are in excellent agreement with the theoretical estimate of \cite{K:96}, given by 0.5264,0.4974 and 0.4686 compared to 0.52,0.46 and 0.46 in the latter. It is noted that, for \Ray$<10^6$,  the $\sigma_{uc}$ increase is steeper than both $\sigma_{ub}$ and $\sigma_{vc}$, while beyond this range their slopes become nearly same. The same observation is made by \citet{K:96} where, however, \Ray=$5 \times 10^6$ is proposed to be the demarcating point.
\par
\citet{K:96} argued that $\lambda_{\theta}$, defined as the peak distance of $<{\theta^{\prime}}^2>$,  scales with $(Nu-1)$. Thus, if $(Nu-1)$ is interpreted as the turbulent contribution to the heat flux by $v^{\prime}$ and $\theta^{\prime}$, then $\lambda_{\theta}$ taken from the $<{\theta^{\prime}}^2>$ profile  relates closely to the fluctuating part of the heat flux. The present calculations show excellent agreement (see Fig. \ref{fig:velTscale}$c$) of $\gamma_T$ (-0.3068 compared to -1/3) with \citet{K:96} and reasonable agreement of $\lambda_u$ (-0.2063 compared to -1/7) with  \citet{CGHKLTWZZ:89}. 

\subsubsection{Skewness and distribution function}
\begin{figure}[!htb]
\centerline{\includegraphics[scale=0.65]{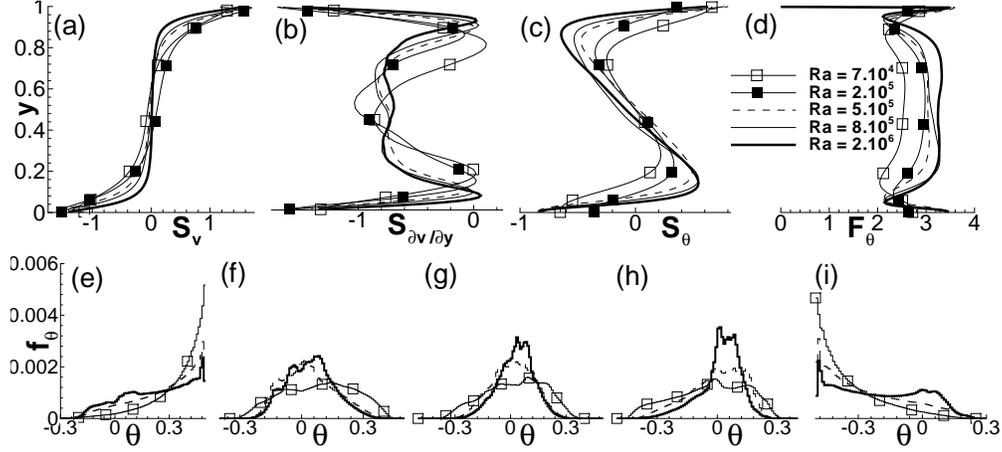}}
\caption{Skewness of $(a)$ $v$ velocity, $(b)$ $\partial v/ \partial y$, $(c)$ temperature, $(d)$ flatness of temperature and $(e-i)$ probability distribution functions of temperature of Regions 1-5. Labels are same as in $(d)$.}\label{fig:skw}
\end{figure}

The skewness of $v$ ($S_v$), its vertical derivative ($S_{\partial v / \partial y}$) and $\theta^{\prime}$  ($S_{\theta}$) along with the temperature flatness ($F_{\theta}$), defined as
\[
S_v=\frac{<v^3>}{{<v^2>}^{3/2}},~S_{\partial v / \partial y}=\frac{<(\frac{\partial v}{\partial y})^3>}{{<(\frac{\partial v}{\partial y})^2>}^{3/2}},~S_{\theta}=\frac{<{\theta^{\prime}}^3>}{{<{\theta^{\prime}}^2>}^{3/2}},~F_{\theta}=\frac{<{\theta^{\prime}}^4>}{<{\theta^{\prime}}^2>^2}
\]
are shown in Fig. \ref{fig:skw}$(a-d)$. 
In situations with a heated lower surface and an insulated top, positive values of $S_v$ were reported by \citet{LWP:80}, \citet{AFB:86} and \citet{MR:90}. In the absence of cold descending plumes, hot plumes ($v > 0$) accelerate near the bottom boundary and dominate the flow locally resulting in the rise of $S_v$ from zero to a positive value through the surface layer. On the other hand, the no-slip Rayleigh-B$\acute{{\bf \mbox{e}}}$nard problem can be thought of two free surface situations placed symmetrically about the center-line, producing two symmetrical curves placed one above the other. Fig. \ref{fig:skw}$(a)$ shows that $S_v$ switches its sign across the central region which is consistent with \cite{WD:74}, \cite{EHZ:86}, \cite{MR:90} and \cite{K:96}. Cold descending plumes that collide with the bottom boundary layer (triggering instabilities that result in eruptions of hot accelerating plumes there) are faster than the newly-formed hot plumes leading to a negative $S_v$ near the bottom plate, and vice versa. The observed fact that $S_v$ is essentially zero in the central region further underlines that fluid acceleration occurs mainly near the plates. The dominance of deceleration of the opposite temperature fluid near the plates is also seen by the negativity of $S_{\partial v / \partial y}$ in Fig. \ref{fig:skw}$(b)$, (as $S_{\partial v / \partial y} < 0$ if hot/cold fluid is decelerating) supporting the predictions of \cite{CGHKLTWZZ:89} and \cite{K:96}. The homogeneity in the central region is strongly supported as, at higher Rayleigh numbers, even these higher-order statistics remain almost invariant in the bulk.
\par
Plumes of opposite temperature coming from the other plate dominate the background flow near both plates,  causing negative temperature variations at the bottom plate and positive variations at the top. This results in the pattern of $S_{\theta}$ seen in Fig. \ref{fig:skw}$(c)$. In the central region, $v^{\prime}$ is symmetrically distributed in both directions  yielding $S_v \approx 0$, while more extreme fluctuations in $\theta$ near the surface layers gradually change across the central region resulting in inhomogeneous behavior of $S_{\theta}$.
\par
The temperature distribution function ($f_{\theta}$) in five distinct volumetric regions, each spanning the entire horizontal extent of the domain, having vertical positions $0 \le y \le 0.1,~0.12 \le y \le 0.3,~0.35 \le y \le 0.65,~0.7 \le y \le 0.88~\mbox{and}~0.9 \le y \le 1$ and denoted by Region 1 to 5, respectively, are shown in Fig. \ref{fig:skw}$(e-i)$. Negatively skewed distributions of $f_{\theta}$ shift to positively skewed ones with long tails as one moves away from the bottom plate, an observation consistent with \cite{CGHKLTWZZ:89}. Beyond the surface layers, asymmetric distributions about a zero mean in Regions 2 and 4 progress to a symmetric profile in the bulk (Region 3). The emergence of a Gaussian profile at higher Rayleigh numbers is further confirmed by the $F_{\theta}$ curve (see Fig. \ref{fig:skw}$d$) which attains a central value of 3.5, closer to the value for a Gaussian ($F_{\theta}=3$) than to an exponential ($F_{\theta}=6$) distribution. It should be noted here that at the same aspect-ratio, \cite{K:96} found $F_{\theta}$ to be nearly 4.5 at \Ray$=2 \times 10^7$. 

\subsubsection{Wave number spectra and Kolmogorov length scale}

Two-dimensional spectra are computed by (a) first carrying out a two-dimensional Fourier transform on instantaneous planar fields with the Fourier coefficient $c_{lm}$ corresponding to the wavenumber set $(\kappa_{xl},\kappa_{zm})$ in $x$ and $z$-directions having $N_x$ and $N_z$ nodal points, respectively,
\[
c_{lm}=\frac{1}{N_x} \frac{1}{N_z} \sum_{p=1}^{N_x} \sum_{q=1}^{N_z} \phi_{pq} \exp [-\hat{i}(\kappa_{xl}i+\kappa_{zm}j) . 
(x_p i+z_q j)]
\]
and (b) summing all power coefficients within unit intervals of $\kappa_h~(\equiv \sqrt{{\kappa_x}^2+{\kappa_z}^2})$ which are (c) then time-averaged. 
\par
\begin{figure}[!htb]
\centerline{\includegraphics[scale=0.62]{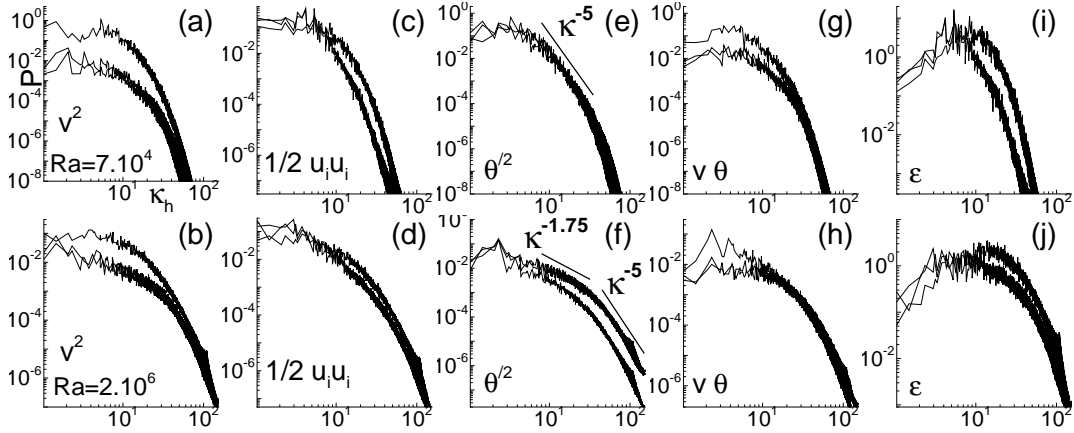}}
\caption{Power spectra at $y=0.1,0.5$ and 0.9; top row: \Ray=$7 \times 10^4$ and bottom row: \Ray=$2 \times 10^6$.}\label{fig:spectra}
\end{figure}

Figure \ref{fig:spectra} $(a-j)$ compares the spectra of the vertical velocity squared $(v^2)$, total kinetic energy $(1/2~u_iu_i)$, temperature fluctuation squared $({\theta^{\prime}}^2)$, vertical heat flux $(v \theta)$ and dissipation of kinetic energy $(\epsilon)$ for the highest and lowest Rayleigh numbers. In all the figures, the separate single curve corresponds to $y=0.5$ while the other two curves (falling on each other) correspond to $y=0.1$ and 0.9. 
In all the spectra, a greater excitation of higher wavenumber modes is evident as Rayleigh number increases. The levels of $v^2$ spectra in the near-wall points are lower to those in the bulk for scales which are not important for dissipation. However, the total kinetic energy is roughly constant across the flow - the increase in the horizontal kinetic energy (caused by circulating flow) in the near-wall regions possibly compensating for the lower vertical kinetic energy there. The $v\theta$ spectra match closely with the $v^2$ spectra, emphasizing the similarity in the behavior of $v$ with $\theta$  across the box at \emph{all} scales. This indicates the pervasive role of buoyancy in shaping the velocity field even while the origin of $\theta^{\prime}$ is not driven solely by the buoyancy acceleration. The peaks of the $\epsilon$ spectra shift to higher $\kappa_h$ at the center $(y=0.5)$, and maximum dissipation occurs at higher wavenumbers ($\kappa_h$) at higher Rayleigh numbers. The centroid of the $\epsilon$ curves shift upwards by at least a decade from the lowest to highest Rayleigh number. 
\par
The ${\theta^{\prime}}^2$ spectra have a nature that is different from the other computed spectra. The large-scale energy levels of ${\theta^{\prime}}^2$ are approximately the same in all regions at all Rayleigh numbers, but near-boundary regions have higher small-scale energy compared to the bulk region at higher Rayleigh numbers (see Fig. \ref{fig:spectra}$f$). At lower Rayleigh numbers, a rapid fall ($\kappa^{-5}$) at lower cut-off wavenumbers is observed. However, at higher Rayleigh numbers, the spectra follow a $\kappa^{-1.75}$ law, close to the $\kappa^{-5/3}$ law for a passive scalar in isotropic turbulence, followed by a steeper $\kappa^{-5}$ law in the dissipative range. This observation is not consistent with the $\kappa^{-1}$ law of \cite{K:96}.
\par
\begin{figure}[!htb]
\centerline{\includegraphics[scale=0.62]{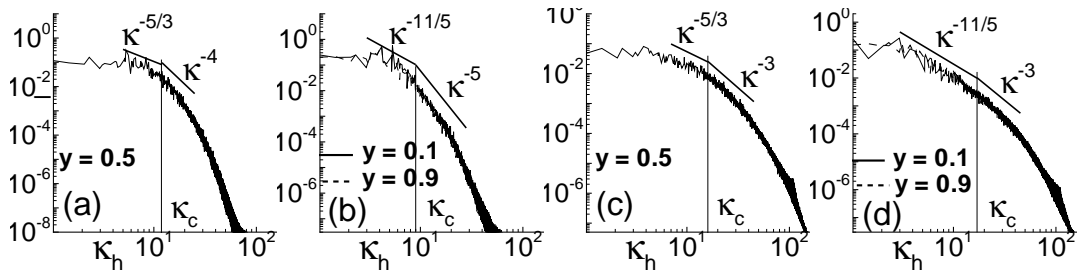}}
\caption{Power-laws for the kinetic energy spectra.}\label{fig:spcpow}
\end{figure}

The power law variation of the kinetic energy spectra is shown in detail in Fig. \ref{fig:spcpow}$(a-d)$ where $\kappa^{-5/3}$ and $\kappa^{-11/5}$ laws at intermediate $\kappa_h$ and a $\kappa^{-p}(p \ge 3)$ decay laws at high $\kappa_h$ are also shown for comparison. 
These two spectral regions shift towards higher wavenumbers with increasing Rayleigh number and the transitional wavenumber between them at \Ray$=2 \times 10^6$ is $\kappa_c \approx 18$. The high $\kappa_h$  exponent $p$ was reported to be 3 for integrated spectra across the box by \citet{K:96} with $\kappa_c=30$ for \Ray=$2 \times 10^7$. Thus, the results are seemingly consistent with the DNS of \cite{K:96} which is the published study closest to the present cases in terms of the geometrical parameters and Rayleigh number range simulated. Theoretically, the high wavenumber part of the spectra should decay at a rate higher than any finite power of $\kappa$ if the velocity field is infinitely differentiable, as discussed by \citet{pope:00}. \citet{kr:59} and \citet{SV:94} showed that exponential decay is realized at sufficiently high Reynolds numbers. Thus, as Reynolds number can be roughly approximated by the square root of Rayleigh number, and the Rayleigh numbers  simulated in the present study are not very high, exponential decay is only partly realized, as expected. It should be mentioned that, in contrast to the large aspect ratio container, a $\kappa^{-5/3}$ law up to the highest wavenumber was reported by \citet{S:88} for all-rigid boundaries simulations. The mixed power-law ($\kappa^{-5/3}$ and $\kappa^{-3}$) obtained here and by \cite{K:96} is consistent with the two-dimensional theoretical predictions of \cite{K:67} and \cite{B:69} for stratified media. They, however, modeled the spectral dynamics as a backward energy cascade and a forward enstrophy cascade with injection of energy and enstrophy at the transitional wavenumber. Further investigations show evidence of a $\kappa^{-11/5}$ variation, not observed by \citet{K:96}, in the near-wall regions (see Fig. \ref{fig:spcpow}$d$). The $\kappa^{-11/5}$ prediction of \cite{B:59} was supported by the velocity structure functions experimentally obtained by \cite{TS:92}. Thus, the present results partly support the shell model of \citet{B:59} which predicted $\kappa^{-11/5}$ for $1/2 u_i u_i$ and $\kappa^{-7/5}$ (compared to -1.75 in the present study) for the ${\theta^{\prime}}^2$ spectra.      
\par
Kolmogorov's hypothesis postulates the existence of an ``inertial" wavenumber region between the energy containing and the dissipative ranges at high Reynolds numbers. This range is characterized by a length scale $l$ such that $\eta \ll l \ll L$, where $\eta$ and $L$ are respectively the Kolmogorov length-scale and the length-scale of the energy-containing range. With two limiting regions being $l \sim \eta$ (or $\kappa\eta \sim 1$) in the dissipative range and $l \sim L$ (or $\kappa\eta \sim \eta/L$) in the energy containing range. Most of the kinetic energy of the flow resides at the lower end in the $\kappa \eta$ scale ($\kappa\eta \sim \eta/L$) while the bulk of the dissipation occurs near the high end of this scale ($\kappa\eta \sim 1$)\cite{LX:10}.  
\par
\begin{figure}[!htb]
\centerline{\includegraphics[height=4cm,width=13cm]{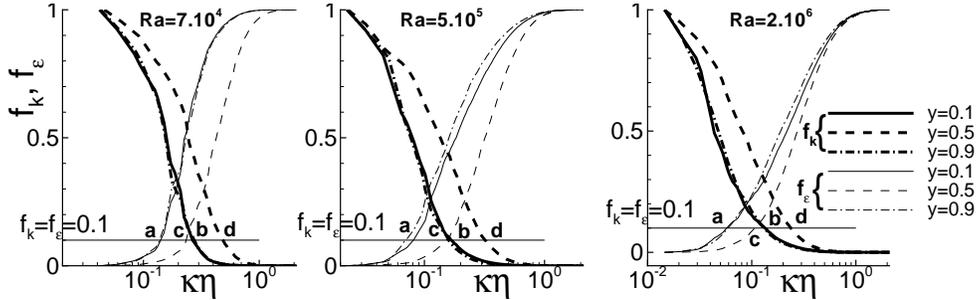}}
\caption{Overlap between the inertial and dissipation ranges.}\label{fig:overlap}
\end{figure}

Figure \ref{fig:overlap} shows $f_k$, the fraction of cumulative kinetic energy \emph{above} a wavenumber $\kappa$, and $f_{\epsilon}$, the dissipation \emph{below} a wavenumber $\kappa$
\[
f_k=\frac{k_{\kappa-\infty}}{k_{0-\infty}}=\frac{\int_{\kappa}^{\infty}k(\kappa) d\kappa}{\int_0^{\infty}k(\kappa) d\kappa},~~~~
f_{\epsilon}=\frac{\epsilon_{0-\kappa}}{\epsilon_{0-\infty}}=\frac{\int_0^{\kappa}\epsilon(\kappa) d\kappa}{\int_0^{\infty}\epsilon(\kappa) d\kappa}
\]  
on the scaled wavenumber ($\kappa\eta$) range. The centroids of the $f_k$ and $f_{\epsilon}$ curves are quite far on the $\kappa \eta$ scale; however, there is some overlap of the two curves. In each plot, a horizontal line corresponding to $10\%$ of the cumulative fraction $(f_k=f_{\epsilon}=0.1)$ is drawn which cuts the $f_k$ and $f_{\epsilon}$ curves at $(b,d)$ and $(a,c)$ points, respectively. As $(\kappa \eta)_b > (\kappa \eta)_a$ and $(\kappa \eta)_d > (\kappa \eta)_c$ the lengths $ab$ and $cd$, denoted by $\Delta_1[\equiv (\kappa \eta)_b-(\kappa \eta)_a]$ and $\Delta_2[\equiv (\kappa \eta)_d-(\kappa \eta)_c]$ are inversely related to the degree of separation of the energy containing and dissipation ranges near the walls and in the central region, respectively. 
\par
Table \ref{kdisp} lists the total integrated kinetic energy and dissipation which appear in the denominator of the ratios $f_k$ and $f_{\epsilon}$, respectively. 
\begin{table}
\begin{center}
\begin{tabular}{ccccccc}\hline
 & & $k_{0-\infty}$ & & & $\epsilon_{0-\infty}$ & \\ 
\Ray & $y=0.1$ & $y=0.5$ & $y=0.9$ & $y=0.1$ & $y=0.5$ & $y=0.9$ \\[3pt]
$7 \times 10^4$   & 1.26759 & 0.96818 & 1.21322 & 28.35227 & 51.04629 & 25.43707 \\ 
$5 \times 10^5$   & 1.01299 & 0.75624 & 1.04588 & 30.60102 & 67.38304 & 27.93405 \\
$2 \times 10^6$   & 0.63506 & 0.50166 & 0.68647 & 29.78119 & 61.23454 & 30.27028 \\ \hline
\end{tabular} 
\caption{Total integrated kinetic energy and dissipation.} 
\label{kdisp}
\end{center}
\end{table}
It can be seen that the total kinetic energy near the walls $(y=0.1,0.9)$ are higher than in the central region while the dissipation in the latter region is around twice of that in the near-wall regions. At all Rayleigh numbers, $k_{\kappa-\infty}$ is higher at $y=0.5$ than at $y=0.1$ or 0.9 clearly indicating a wider range of energy containing eddies in the central region which can be attributed to strong mixing and chaotic interaction leading to the distribution of turbulent kinetic energy over a wider range of length scales. On the other hand, $\epsilon_{0-\kappa}$ is higher in the near-wall regions compared to the bulk as the wider spread of turbulent kinetic energy in the central region leads to a greater fraction of dissipation at higher wavenumbers.  
\par
\begin{figure}[!htb]
\centerline{\includegraphics[height=5.5cm,width=12cm]{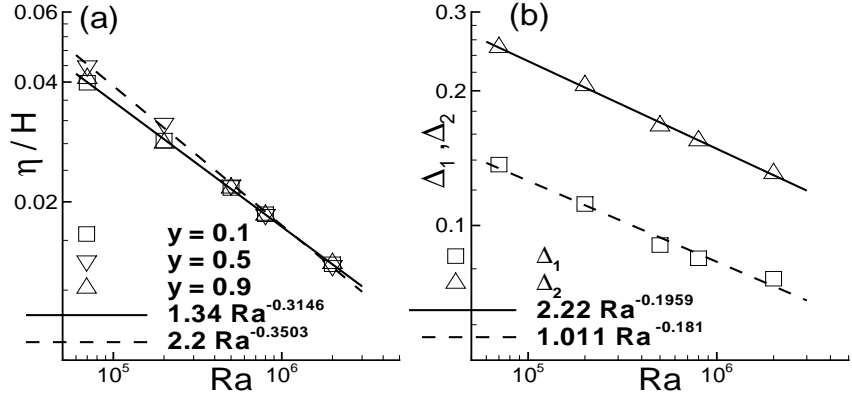}}
\caption{Power-laws for the $(a)$ Kolmogorov length scale and $(b)$ separation scales $\Delta_1,~\Delta_2$.}\label{fig:etainer}
\end{figure}

From the curves in Fig. \ref{fig:overlap}, $\kappa \eta$ values corresponding to the points $a,b,c,d$ are shown in table \ref{delta}. 
\begin{table}
\begin{center}
\begin{tabular}{cccc} \hline
\Ray & $ \Delta_1~(y=0.1)$ & $\Delta_1~(y=0.9)$ & $\Delta_2$ \\[3pt]
$7 \times 10^4$   & 0.13210 & 0.14148 & 0.25000 \\ 
$5 \times 10^5$   & 0.09268 & 0.08814 & 0.16699 \\
$2 \times 10^6$   & 0.08025 & 0.07179 & 0.13067 \\ \hline
\end{tabular} 
\caption{Separation in $\kappa \eta$ scale as in Fig. \ref{fig:overlap}.} 
\label{delta}
\end{center}
\end{table}
It is observed that $\Delta_1 < \Delta_2$ at all Rayleigh numbers, indicating a greater separation (or lesser overlap) of the energy containing and dissipative ranges in the near-wall regions compared to the bulk. Moreover, both $\Delta_1$ and $\Delta_2$ consistently decrease with increase in Rayleigh numbers --  which is expected -- as at higher Rayleigh numbers these ranges would move further away in the $\kappa$ space indicating an increasingly larger inertial subrange. The power-law variation is shown in Fig. \ref{fig:etainer}$(a)$ where $\Delta$ values are averaged near the two walls to determine a single scaling law. They are found to be
\[
\Delta_1=1.011~\Ray^{-0.181},~~~~~\Delta_2=2.22~\Ray^{-0.1959}
\] 
The decay exponents for the near-wall and the central regions turn out to be almost the same.  
\par
If the dissipation is normalized by $U^3/H~(U \equiv  \sqrt{g\beta\Delta TH})$, the smallest scale $(\nu^3/\epsilon)$ becomes 
\begin{equation}
\label{etadis}
\frac{\eta}{H}=\left(\frac{\Pran}{\Ray}\right)^{3/8} {\epsilon^{*}}^{-1/4},
~~\mbox{with}~\epsilon^{*} \equiv \int_0^{\infty} \epsilon^{*}(\kappa_h) d\kappa_h
\end{equation}      
where $\epsilon^{*}(\kappa_h)$ is the dissipation spectrum at the horizontal wavenumber $\kappa_h~(\equiv \sqrt{{\kappa_x}^2+{\kappa_z}^2})$ computed from non-dimensional solutions, shown in Fig. \ref{fig:spectra}$(i,j)$. The Kolmogorov length scale computed using \eqref{etadis} and that of \cite{K:96} calculated from  
\begin{equation}
\label{etaNu}
\frac{\eta}{H}=\left( \frac{{\Pran}^2}{(Nu-1)\Ray} \right)^{1/4}
\end{equation}
(which assumes that normalized dissipation equals $(Nu-1)\Ray$) are shown in table \ref{eta}. The present $\eta$ values are consistent with \cite{K:96} with central values being closest to the reference. 
\begin{table}
\begin{center}
\scalebox{0.8}{
\begin{tabular}{ccccc}\hline
\Ray & & Present $\eta/H$ & & $\eta/H$ by \cite{K:96} \\
 & $y=0.1$ & $y=0.5$ & $y=0.9$ \\[3pt] 
$7 \times 10^4$   &  0.0398375400 & 0.0438951542 &  0.0408495706 & 0.05 ($5\times 10^4$) \\ 
$2 \times 10^5$   &  0.0285082552 & 0.0314740218 &  0.0281044211 & 0.0325 ($2\times 10^5$)\\ 
$5 \times 10^5$   &  0.0216694393 & 0.0220269361 &  0.0216723746 & 0.024 ($5\times 10^5$)\\ 
$8 \times 10^5$   &  0.0186121644 & 0.0185780961 &  0.0185226249 &  \\ 
$2 \times 10^6$   &  0.0139296194 & 0.0138163561 &  0.0140018226 & 0.014 ($2.5\times 10^6$)\\ \hline
\end{tabular}}
\caption{Kolmogorov length scale at $y=0.1,0.9$ and 0.5 using Eq.~\eqref{etadis} and Eq.~\eqref{etaNu}.}
\label{eta} 
\end{center}
\end{table}
Near the walls, $\eta$ is only slightly smaller than in the central region, with the difference reducing  with increase in Rayleigh number, so that at \Ray=$2\times 10^6$ almost the same $\eta$ values are obtained everywhere. Fig. \ref{fig:etainer}$(b)$ shows the least-square power-law curves for the computed Kolmogorov length scales: $\eta/H= 1.337~\Ray^{-0.3146}$ at $y=$0.1,0.9 and $\eta/H= 2.207~\Ray^{-0.3503}$ at $y=$0.5  which are in agreement with \cite{K:96}, $\eta/H\approx 1.3~\Ray^{-0.32}$. Thus \eqref{etadis} and \eqref{etaNu} are close to equivalent and the validity of the $\epsilon^{*}=(Nu-1)\Ray$ approximation is supported here.

\subsubsection{Production and dissipation of turbulent kinetic energy and enstrophy}
The dynamical equations for the turbulent kinetic energy $(1/2~\overline{u_i^{\prime}u_i^{\prime}})$ and turbulent enstrophy ($1/2~\overline{\omega_i^{\prime}\omega_i^{\prime}}$) are (\cite{TL:72}): 
\[
\underbrace{\frac{\partial}{\partial t} (\frac{1}{2}\overline{u_i^{\prime}u_i^{\prime}}) + 
U_j\frac{\partial}{\partial x_j} (\frac{1}{2}\overline{u_i^{\prime}u_i^{\prime}})}_{A_k} = 
-\underbrace{\frac{\partial}{\partial x_j} \left (\overline{p^{\prime}u_j^{\prime}} - 2  
(\Pran/\Ray)^{\frac{1}{2}} ~\overline{u_i^{\prime}s_{ij}^{\prime}}+
\frac{1}{2} \overline{u_i^{\prime}u_i^{\prime}u_j^{\prime}} \right )}_{T_k}
\]
\begin{equation}
\label{tke}
+\underbrace{\overline{u_i^{\prime}\theta^{\prime}}~\delta_{iy}}_{P_B} 
- \underbrace{\overline{u_i^{\prime}u_j^{\prime}}~S_{ij}}_{P_S} 
-\underbrace{2 (\Pran/\Ray)^{\frac{1}{2}} ~\overline{s_{ij}^{\prime}s_{ij}^{\prime}}}_{\epsilon}
\end{equation} 
and
\[
\underbrace{\frac{\partial}{\partial t} (\frac{1}{2}\overline{\omega_i^{\prime}\omega_i^{\prime}}) + 
U_j\frac{\partial}{\partial x_j} (\frac{1}{2}\overline{\omega_i^{\prime}\omega_i^{\prime}})}_{A_E} = 
-\underbrace{\overline{u_j^{\prime}\omega_i^{\prime}} ~\frac{\partial \Omega_i}{\partial x_j}}_{P_{EG}}
-\underbrace{\frac{\partial}{\partial x_j} (\frac{1}{2} \overline{\omega_i^{\prime}\omega_i^{\prime}u_j^{\prime}})}_{T_E}
\]
\[
+\underbrace{\Omega_j ~\overline{\omega_i^{\prime}s_{ij}^{\prime}}}_{P_{EM}}
+\underbrace{\overline{\omega_i^{\prime}\omega_j^{\prime}s_{ij}^{\prime}}}_{P_{ETS}}
+\underbrace{\overline{\omega_i^{\prime}\omega_j^{\prime}} ~S_{ij}}_{P_{EMS}}
\]
\begin{equation}
\label{enstr}
+\underbrace{(\Pran/\Ray)^{\frac{1}{2}} ~\frac{\partial^2}{\partial x_j \partial x_j} (\frac{1}{2}\overline{\omega_i^{\prime}\omega_i^{\prime}})}_{T_v}
-\underbrace{(\Pran/\Ray)^{\frac{1}{2}} ~\overline{\frac{\partial \omega_i^{\prime}}{\partial x_j} ~\frac{\partial \omega_i^{\prime}}{\partial x_j}}}_{\epsilon_E}
+\underbrace{\Pi_{ijk} ~\overline{\omega_i^{\prime}\frac{\partial \theta^{\prime}}{\partial x_j}} 
~\delta_{ky}}_{P_{EB}}
\end{equation}
where $u_i^{\prime}=u_i-U_i,~s_{ij}^{\prime}=s_{ij}-S_{ij},~p^{\prime}=p-P,~\theta^{\prime}=\theta-\Theta, ~{\omega_i}^{\prime}~=\omega_i-\Omega_i,~\delta_{ij}$ is the kronecker delta (1 if $i=j$ and 0 otherwise) and $\Pi_{ijk}$ is the alternating tensor (1 if $i,j,k$ cyclic, -1 if anti-cyclic and 0 otherwise). Both the equations contain advection ($A_k,A_E)$, production ($P_B,P_S,P_{EG},P_{EM},P_{ETS},P_{EMS},P_{EB}$), transport ($T_k,T_E,T_v$) and dissipation ($\epsilon,\epsilon_E$) terms, given in detail by \cite{TL:72}. Unlike in shear flows, where $P_B$ is absent and the shear production  $\overline{u_i^{\prime}u_j^{\prime}}S_{ij}$ bleeds energy from the mean flow and feeds the turbulence, the buoyancy production is not directly linked to the mean flow kinetic energy and owes its origin purely due to the correlation between $v^{\prime}$ and $\theta^{\prime}$. 
\par
\begin{figure}[!htb]
\centerline{\includegraphics[height=5cm,width=11cm]{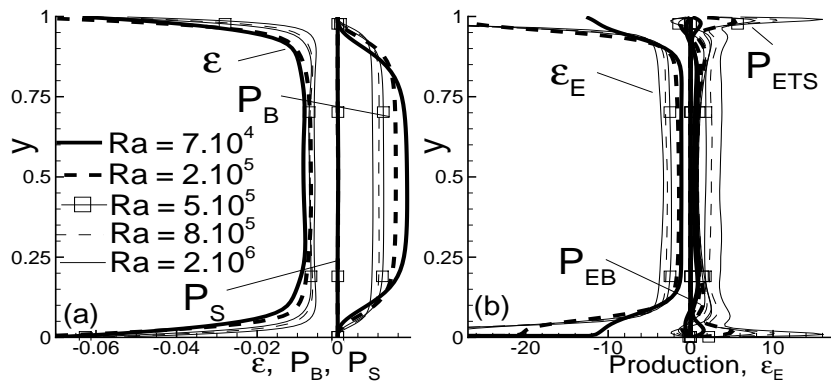}}
\caption{Productions and dissipation of the $(a)$ turbulent kinetic energy and $(b)$ turbulent enstrophy, labels are same as in $(a)$.}\label{fig:ke}
\end{figure}

Figure \ref{fig:ke}$(a)$ shows the vertical profiles (time averages of the $y$-planar averages over several integral time scales) of the productions and dissipation of turbulent kinetic energy. 

It is observed that shear does not contribute to the production of turbulent kinetic energy as $-\overline{u_i^{\prime}u_j^{\prime}}~S_{ij}$ is at least 3-orders smaller than $\overline{v^{\prime}\theta^{\prime}}$. The buoyancy production $P_B$ is zero at the walls, and increases with a steep gradient to reach a high value by the edge of the boundary layer while the dissipation drops from its maximum to an asymptotic value across the boundary layer. In the vicinity of walls, though high dissipation is observed, buoyancy has little effect in producing kinetic energy -- which points out the importance of viscous transport in that region, as was observed in boundary layers by \citet{KMM:87}. Both $\overline{v^{\prime}\theta^{\prime}}$ and $\epsilon$ have relatively constant values outside the boundary layers (which become thinner with increase in Rayleigh numbers), and an approximate balance between them is found outside the surface layer, i.e.,  
\begin{equation}
\label{keb}
\overline{v^{\prime}\theta^{\prime}} \approx \nu ~\overline{s_{ij}^{\prime} s_{ij}^{\prime}}
\end{equation}
This is also evident from table \ref{balance} where the production and dissipation terms integrated over the whole domain are shown. 
\begin{table}
\begin{center}
\scalebox{0.8}{
\begin{tabular}{cccc}\hline
\Ray & $\int P_S ~dy$ & $\int P_B ~dy$ & $\int \epsilon ~dy$ \\[3pt]
$7 \times 10^4$   &   $2.389~10^{-5}$   &   $1.283~10^{-2}$   &   $-1.251~10^{-2}$ \\ 
$2 \times 10^5$   &   $1.844~10^{-5}$   &   $1.171~10^{-2}$   &   $-1.127~10^{-2}$ \\ 
$5 \times 10^5$   &   $1.361~10^{-5}$   &   $9.725~10^{-3}$   &   $-9.266~10^{-3}$ \\ 
$8 \times 10^5$   &   $1.834~10^{-5}$   &   $8.994~10^{-3}$   &   $-8.619~10^{-3}$ \\ 
$2 \times 10^6$   &   $3.745~10^{-6}$   &   $7.761~10^{-3}$   &   $-7.138~10^{-3}$ \\ \hline
\end{tabular}}
\caption{Integrated production and dissipation showing the balance \eqref{keb}, $P_B \approx \epsilon$}
\label{balance} 
\end{center}
\end{table}
As shear production is very small, either Reynolds stresses and mean shear have essentially zero correlation with each other or the principal mean strain rate is zero. From an examination of the instantaneous planar-averaged values of $P_S$ (which are non-zero) it is evident that the former is true (in this problem the time-averaged value of $S_{ij}$ is zero, due to homogeneity, however the $S_{ij}$ here is the instantaneous planar average, which is multiplied by the instantaneous planar average Reynolds stress $\overline{{u_i}^{\prime}{u_j}^{\prime}}$ and \emph{then} time-averaged. This final quantity need \emph{not} be zero from considerations of homogeneity alone. It should be mentioned that in this study, planar averages are treated essentially as ensembles-averages, when applied to turbulent quantities). 
\par
In the presence of a negative vertical mean temperature gradient ($\partial \Theta/\partial y<0$),  $\overline{v^{\prime}\theta^{\prime}}>0$ causes upward heat transfer. The $P_B \approx \epsilon$ balance indicates $v^{\prime}$ and $\theta^{\prime}$ are tuned to the same band of frequencies and they sustain turbulence and vice versa at the bulk of the flow, indicating that buoyancy effects are pervasive across all scales. Unlike the kinetic energy budget in shear flows, Eq.~\eqref{keb} indicates that interaction between the mean flow and the fluctuating field is not responsible for the sustenance of turbulence in turbulent Rayleigh-B$\acute{{\bf \mbox{e}}}$nard convection.
\par
There are five sources of enstrophy production in \eqref{enstr} involving both mean and fluctuating fields. The gradient production ($P_{EG}$) and mixed production ($P_{EM}$) due to the mean vorticity and mean shear ($P_{EMS}$) involve interactions between the mean and turbulent fields, while the production due to turbulent stretching ($P_{ETS}$) and buoyancy ($P_{EB}$) require correlations among fluctuating fields. \citet{TL:72} argued that for shear flows production due to turbulent stretching ($\overline{\omega_i^{\prime}\omega_j^{\prime}s_{ij}^{\prime}}$) is higher than all the other terms (except dissipation) by at least a factor of $O(Re^{1/2})$ at sufficiently high Reynolds numbers. The approximate enstrophy budget in shear flow turbulence was provided by \citet{T:38} as
\begin{equation}
\label{enstrsh}
\overline{\omega_i^{\prime}\omega_j^{\prime}s_{ij}^{\prime}}=\nu ~\overline{\frac{\partial \omega_i^{\prime}}{\partial x_j}\frac{\partial \omega_i^{\prime}}{\partial x_j}}    
\end{equation}
Figure \ref{fig:ke}$(b)$ shows time-averages of various production and dissipation terms of Eq.~\eqref{enstr}. It is to be noted that $P_{EG},P_{EM}$ and $P_{EMS}$, all associated with the mean vorticity and mean shear fields, are essentially zero and that outside the surface layers an approximate balance between production due to turbulent stretching $(P_{ETS})$, buoyancy $(P_{EB})$ and dissipation $(\epsilon_E)$ exists, i.e.,
\begin{equation}
\label{enstrb}
\overline{\omega_i^{\prime}\omega_j^{\prime}s_{ij}^{\prime}}+
\left( ~\overline{\omega_z^{\prime} \frac{\partial \theta^{\prime}}{\partial x}}-
\overline{\omega_x^{\prime} \frac{\partial \theta^{\prime}}{\partial z}} ~\right) \approx \nu ~\overline{\frac{\partial \omega_i^{\prime}}{\partial x_j}\frac{\partial \omega_i^{\prime}}{\partial x_j}}    
\end{equation}   
This result modifies the classical expression Eq.~\eqref{enstrsh} for shear flow turbulence by including the effect of buoyancy.
\par
\begin{figure}[!htb]
\centerline{\includegraphics[height=5.5cm,width=12cm]{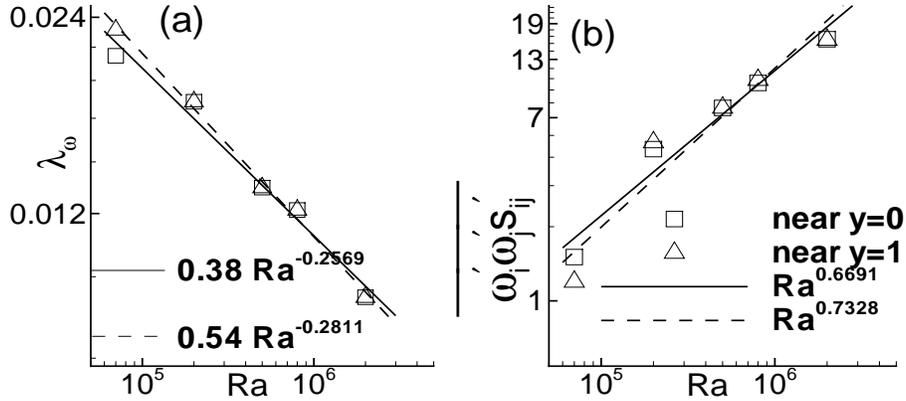}}
\caption{Power-laws for $(a)$ peak distance and $(b)$ peak values of the turbulent stretching $\overline{\omega_i^{\prime}\omega_j^{\prime}s_{ij}^{\prime}}$; symbols correspond to same $y$ in $(a)$ and $(b)$.}\label{fig:enstr_scale}
\end{figure}

The distance $(\lambda_{\omega})$ and magnitude of the peaks of the turbulent stretching ($\overline{\omega_i^{\prime}\omega_j^{\prime}s_{ij}^{\prime}}$) term in the boundary layers are shown in Fig. \ref{fig:enstr_scale}$(a,b)$ along with a least-square fit. 
For $\lambda_{\omega}$, the two lines corresponding to the top and bottom boundary layers nearly fall on each other, with power-law exponents being $-0.2569$ and $-0.2811$ which lie between the exponents of $\lambda_{\theta}$ and $\lambda_u$ of Fig. \ref{fig:velTscale}$(c)$. This indicates that strain rate fluctuations produce the maximum mean squared vorticity fluctuations in the  boundary layer, where horizontal velocities and temperature oscillations also reach their maximum. The exponents of the peak values of $P_{ETS}$, 0.6691 and 0.7328, are found to be higher than $\sigma_{ub}$ in Fig. \ref{fig:velTscale}$(b)$. However, the peak $P_{ETS}$ magnitude increases more steeply with Rayleigh number than the peaks of $\sigma_{ub}$ and $\sigma_{uc}$.      

\subsection{$<Nu>$-\Ray ~relationship}
\begin{figure}[!htb]
\centerline{\includegraphics[height=7.5cm,width=8.5cm]{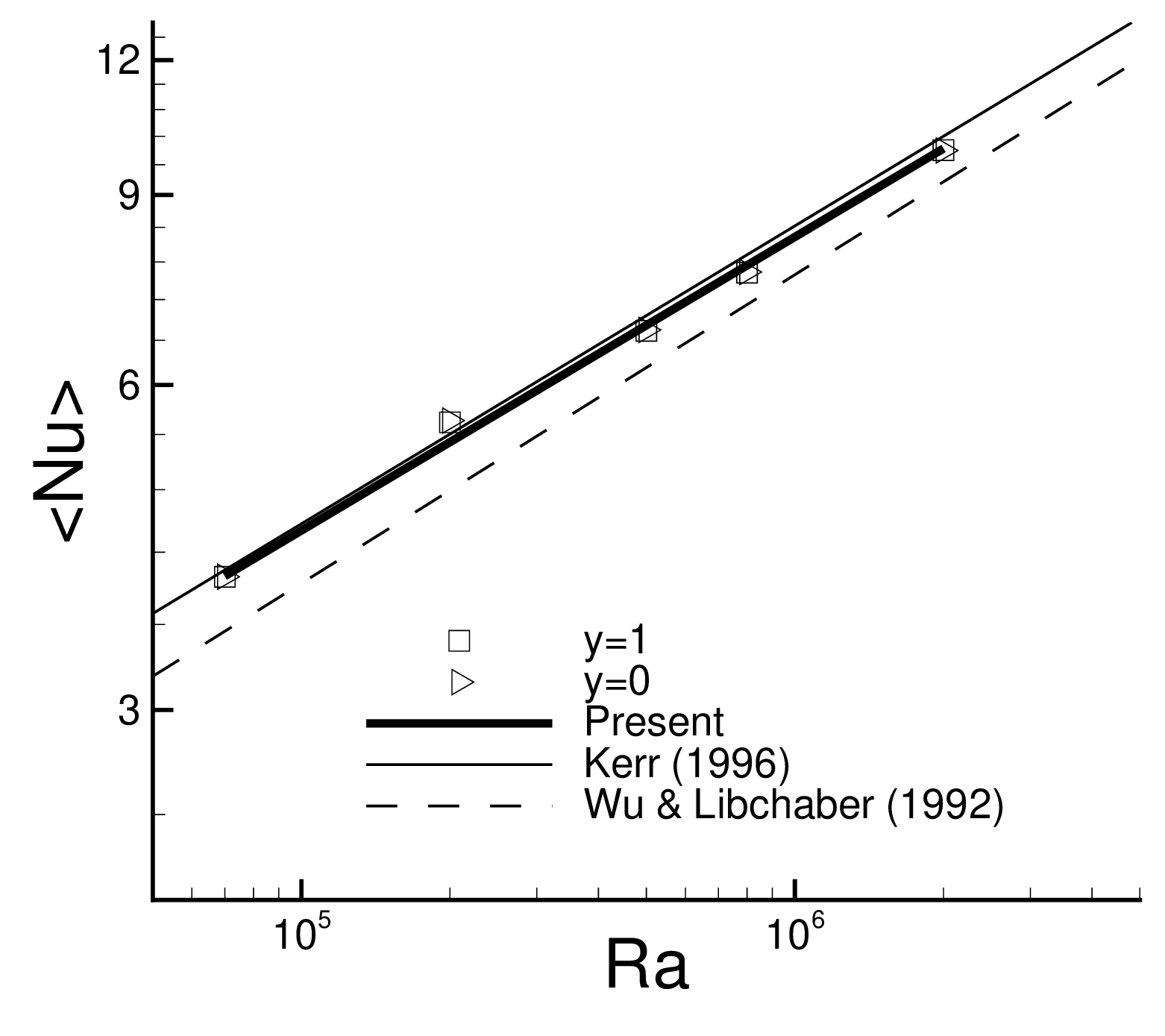}}
\caption{$<Nu>-$ \Ray ~relationship.}\label{fig:nura}
\end{figure}

The departure of $<Nu>-\Ray$ scaling from the theoretical prediction ($Nu \sim \Ray^{1/3}$) has been examined in Fig. \ref{fig:nura}. 
The experiments of \citet{CG:73} established this deviation from the classical theory, which was then taken as a basis for the classification of ``soft" and ``hard" turbulence regimes $(4\times 10^7 < \Ray < 10^{12})$. Since then a number of experiments and numerical studies have reported the Rayleigh number exponent to be close to $2/7(=0.286)$\cite{AGL:09,SPGL:13}. A least-square fit of the present data yields an exponent of 0.272. For comparison the predictions of \citet{WL:92} and \citet{K:96} are also shown in the figure. The present $<Nu>$ curve falls close to the earlier DNS computations of \cite{K:96} where the Rayleigh number extends to $2 \times 10^7$ and the the exponent (0.272) is very close to the value of 2/7 of \cite{WL:92} obtained experimentally for a larger range of Rayleigh number. Table \ref{nuratab} lists results from a few selected studies with their predicted $<Nu>$ exponent. 
\begin{table}
\begin{center}
\scalebox{0.8}{
\begin{tabular}{cccc}\hline
References (range of \Ray) & $A$ & Fluid & $n$ \\ 
Kerr (1996) ($5 \times 10^4 < \Ray < 2 \times 10^7$) & 6 & Air & 0.276 \\
Emran and Schumacher (2012) ($1 \times 10^7 < \Ray < 3 \times 10^{10}$) & 1 & Air & 0.3 \\
Chu \& Goldstein (1973) ($2.76 \times 10^5 < \Ray < 1.05 \times 10^8$) & 6.02 & Water & 0.278 \\ 
Christie \& Domaradzki (1993) ($2.5 \times 10^5 < \Ray < 6.3 \times 10^5$) & 3-6 & Air & 0.274 \\  
Solomon \& Gollub (1990,1991) ($1 \times 10^6 < \Ray < 2 \times 10^8$) & 0.7-1.6 & Water & 0.284 \\ 
Heslot {\it et al.} (1987) ($2.5 \times 10^5 < \Ray < 4 \times 10^7$) & 1 & Helium & 1/3 \\ 
DeLuca {\it et al.} (1990) ($1.28 \times 10^6 < \Ray < 5.12 \times 10^6$) & 1(2D) & \Pran=1 & 1/3 \\
($5.12 \times 10^6 < \Ray < 1.6 \times 10^8$) & & & 2/7 \\ 
Present ($7 \times 10^4 < \Ray < 2 \times 10^6$) & 6 & Air & 0.272 \\ \hline
\end{tabular}}
\caption{$<Nu>=C~\Ray^n$ scaling with aspect ratio ($A$), Rayleigh number range, fluid used in turbulent Rayleigh-B$\acute{{\bf \mbox{e}}}$nard convection literature.}
\label{nuratab}
\end{center}
\end{table} 
It is evident that the exponent obtained from the present simulations is very close to those obtained in experiments and simulations, although the Rayleigh numbers simulated in the present work are much lower than what has been attributed to the ``hard" turbulence regime in experimental studies. Therefore, the $<Nu>$ scaling should be seen only as a yet unexplained departure from the classical theory, and its use in the classification of thermal convection in terms of ``hard" and ``soft" turbulent regimes is unsupported by the present results. 

\section{Conclusions}
Direct numerical simulations of turbulent Rayleigh-B$\acute{{\bf \mbox{e}}}$nard convection using a higher-order finite difference scheme are carried out for $7 \times 10^4 \le \Ray \le 2 \times 10^6$. The primary source of plume formation are boundary layer instabilities caused by the collision of plumes with the wall boundary layers. Horizontal velocities have higher integral time scale near the walls, and are not correlated at different points of the domain with the temperature. On the other hand, the vertical velocity has higher integral time-scales in the bulk region and shows a strong correlation with temperature at all scales. Homogeneity is evident in the bulk region with temperature signals being white-noise-like. 
\par
The presence of a $-5/3$ law in the frequency spectra is seen for all the velocities and the temperature. However, while variances of the horizontal velocities show a single exponent power-law relationship with Rayleigh number, the vertical velocity and temperature show variations of behavior near the walls and in the bulk region. The probability density functions of velocities and temperature show a trend towards Gaussian distributions with increasing Rayleigh number. However, the distributions are too unclear to classify turbulent convection into different regimes of ``soft" and ``hard", at least at the Rayleigh numbers studied. 
\par                          
The vertical velocity skewness, in contrast to free-slip flows, is found to be negative near the bottom wall  and essentially zero in the central region. The turbulent kinetic energy spectra shows mixed power-laws: $\kappa^{-5/3}$ (and $\kappa^{-n},n \ge 3$, beyond a transitional wavenumber) near the center and $\kappa^{-11/5}$ (and $\kappa^{-n},n \ge 3$) near the walls with the transitional wavenumber matching closely with the previous DNS calculations. 
\par
The turbulent kinetic energy budget yields an approximate balance between the buoyancy production and the dissipation, with the production due to shear being negligible. The enstrophy budget for the turbulent shear flows is modified in buoyancy-driven turbulence to a balance between the production due to the turbulent stretching, the production due to the buoyancy and the enstrophy dissipation. 
\par
The $<Nu>-\Ray$ power-law gives an exponent of 0.272, identical to the 2/7 value for ``hard" turbulence. However, as the probability density function of velocity and temperature do not show a clear transition, to purportedly ``hard" turbulence features, it is evident that the ``soft" and ``hard" regimes of turbulent convection need to be more clearly characterized. 

\baselineskip  12pt

\bibliography{refer}

\end{document}